\documentclass{article}
\usepackage{amstex}
\usepackage{epsfig}

\begin{document}

\begin{center}
{\bf\Large The STRESS Method for Boundary-point Performance Analysis of 
End-to-end Multicast Timer-Suppression Mechanisms}

Ahmed Helmy+, Sandeep Gupta+, Deborah Estrin*

+ University of Southern California (USC), * University of California at 
Los Angeles (UCLA)


\end{center}



\begin{abstract}

	The advent of multicast and the growth and complexity of the
Internet has complicated network protocol design and evaluation.
Evaluation of Internet protocols usually uses random scenarios or
scenarios based on designers' intuition. Such approach may be useful for
average-case analysis but does not cover {\em boundary-point} (worst or
best-case) scenarios. To synthesize boundary-point scenarios a more
systematic approach is needed.

	In this paper, we present a method for automatic synthesis of
worst and best case scenarios for protocol boundary-point evaluation.
	Our method uses a fault-oriented test generation (FOTG) algorithm
for searching the protocol and system state space to synthesize these
scenarios.  The algorithm is based on a global finite state machine (FSM)
model. We extend the algorithm with timing semantics to handle end-to-end
delays and address performance criteria. We introduce the notion of a {\em
virtual LAN} to represent delays of the underlying multicast distribution
tree. The algorithms used in our method utilize implicit backward search using branch
and bound techniques and start from given {\em target events}. This aims to reduce the
search complexity drastically.

As a case study, we use our method to evaluate variants of the timer
suppression mechanism, used in various multicast protocols, with respect
to two performance criteria: overhead of response messages and response
time.  Simulation results for reliable multicast protocols show that our
method provides a scalable way for synthesizing worst-case scenarios
automatically. Results obtained using stress scenarios differ dramatically 
from those obtained through average-case analyses. We hope for our method 
to serve as a model for applying systematic scenario generation to other 
multicast protocols. 

\end{abstract}



\section{Introduction}
\label{introduction}

\small

	The longevity and power of Internet technologies derives
from its ability to operate under a wide range of operating
conditions (underlying topologies and transmission 
characteristics, as well as heterogeneous applications generating
varied traffic inputs).
Perhaps more than any other technology, the range of operating
conditions is enormous (it is the cross product of the top and
bottom of the IP protocol stack).

Perhaps it is this enormous set of conditions that has inhibited the
development of systematic approaches to analyzing Internet protocol
designs. How can we test correctness or characterize performance of a
protocol when the set of inputs is intractable?  Nevertheless, networking
infrastructure is increasingly critical and there is enormous need to
increase the understanding and robustness of network protocols. Most
current approaches for protocol evaluation use average-case analysis and
are based on random or intuitive scenarios. Such approach does not address
protocol robustness or {\em boundary-point} analysis, in which the
protocol exhibits worst or best-case behavior. We believe that such
protocol breaking points should be identified and studied in depth to
understand and hopefully increase protocol robustness. It is time to
develop techniques for systematic testing of protocol behavior, even in
the face of the above challenges and obstacles. At the same time we do not
expect that complex adaptive protocols will be automatically verifiable
under their full range of conditions. Rather, we are proposing a framework
in which a protocol designer can follow a set of systematic steps,
assisted by automation where possible, to cover a specific part of the
design and operating space. Our goal is to complement average case studies
and enrich the evaluation test-suites for multicast protocols.

In our proposed framework, a protocol designer will still need to create
the initial mechanisms, describe it in the form of a finite state
machine, and identify the performance criteria or correctness
conditions that need to be investigated.
Our automated method will pick up at that point, providing
algorithms that generate scenarios or test suites
that stress the protocol with respect to the identified criteria.
The algorithms used in our method utilize implicit backward search using branch
and bound techniques and start from given {\em target events}. This aims to reduce the
search complexity drastically, as we shall discuss based on our case studies.

This paper demonstrates our progress in realizing this vision as we present
our method and apply it to {\em boundary-point} (worst and best-case) performance
evaluation of the timer suppression mechanism used  in numerous multicast protocols.

\subsection{Motivation}

	The recent growth of the Internet and its increased
heterogeneity
has introduced new failure modes and added complexity to protocol design and
testing.
In addition, the advent of multicast applications
has introduced new challenges of qualitatively different nature
than the traditional point-to-point protocols. Multicast
applications involve a group of receivers and one or more
senders.
	As more complex multicast applications and protocols are
coming to life, the need for systematic and automatic methods to
study and evaluate such protocols is becoming more apparent. Such
methods aim to expedite the protocol development cycle and
improve resulting protocol robustness and performance.

	Through our proposed methodology for test synthesis, we
hope to address the following key issues of protocol design and
evaluation.

\begin{itemize}
\item Scenario dependent evaluation, and the
use of validation test suites:
Protocols may be evaluated for correctness and performance.
	In many evaluation studies of multicast protocols,
the results are dependent upon several factors,
such as membership distribution and network topology, among others.
Hence, conclusions drawn from these studies 
depend heavily upon the evaluation scenarios.

	Protocol development usually passes through
iterative cycles of refinement, which requires revisiting 
the evaluation scenarios to ensure that no
erroneous behavior has been introduced. 
This brings about the need for validation test suites.
Constructing these test suites can be an onerous and error-prone
task if performed manually. Unfortunately, little work has been
done to automate the generation of such tests for multicast network
protocols.
	In this paper, we propose a method for synthesizing test
scenarios automatically for boundary-point analysis of timer-suppression mechanisms
employed by several classes of Internet multicast protocols.

\item Worst-case analysis of protocols:
	It is difficult to design a protocol that
would perform well in all environments. However, identifying
breaking points that violate correctness or
exhibit worst-case performance behaviors of a protocol may give 
insight to protocol designers and help in evaluating design
trade-offs.
	In general, it is desirable to identify, early on 
in the protocol development cycle, scenarios under which the protocol 
exhibits worst or best case behavior.
The method presented in this paper automates the generation of
scenarios in which multicast protocols exhibit worst and best case 
behaviors.

\item Performance benchmarking:
New protocols may propose to refine a
mechanism with respect to a particular performance metric, using
for evaluation those scenarios that show performance improvement.
However, without systematic evaluation, these refinement
studies often (though unintentionally) overlook other scenarios that may 
be relevant. To alleviate such a problem we propose to integrate
stress test scenarios that provide an objective benchmark for performance 
evaluation.

	Using our scenario synthesis methodology we hope to
contribute to the understanding of better performance benchmarking 
and the design of more robust protocols.

\end{itemize}

\subsection{Background}
\label{building_blocks}

	The design of multicast protocols has introduced new
challenges and problems. Some of the problems are common to a
wide range of protocols and applications. One such problem is the
{\em multi-responder} problem, where multiple members of a group
may respond (almost) simultaneously to an event, which may cause
a flood of messages throughout the network, and in turn may lead
to synchronized responses, and may cause additional overhead (e.g., the well-known 
{\em Ack implosion} problem), leading to performance degradation.

	One common technique to alleviate the above problem is
the {\em multicast damping} technique, which employs a {\em timer
suppression} mechanism (TSM). TSM is employed in several multicast
protocols, including the following:
\begin{itemize}
\item IP-multicast protocols, e.g., PIM~\cite{PIM-SMv2-Spec}~\cite{PIM-DM-SPEC}
and IGMP~\cite{igmp}, use TSM on LANs to reduce Join/Prune control
overhead.
\item Reliable multicast schemes, e.g., SRM~\cite{SRM} and MFTP~\cite{mftp}, 
use this mechanism to alleviate {\em Ack implosion}.
	Variants of the SRM timers are used in registry
replication (e.g., RRM~\cite{rrm}~\cite{rrm2}) and adaptive web
caching~\cite{awc}.
\item Multicast address allocation schemes, e.g.,
AAP and SDr~\cite{sdr}, use TSM 
to avoid an implosion of responses during the collision detection phase.
\item Active services~\cite{elan} use multicast damping
to launch one service agent `servent' from a pool of servers.
\end{itemize}
TSM is also used in self-organizing hierarchies (SCAN~\cite{scan}), and
transport protocols (e.g., XTP~\cite{xtp} and RTP~\cite{RTP}).

We believe TSM is a good building block to analyze as our first end-to-end case study,
since it is rich in multicast and timing semantics, and can be evaluated using
standard performance criteria.
As a case study, we examine its worst and best case behaviors
in a systematic, automatic fashion\footnote{Such behavior is not protocol specific,
and if a protocol is composed of previously checked building blocks, these parts of
the protocol need not be revalidated in full. However, interaction 
between the building blocks still needs to be validated.}.

In TSM, a member of a multicast group that has detected
loss of a data packet multicasts a request for recovery. Other
members of the group, that receive this request and that have previously received the
data packet, schedule transmission of a response. In general, randomized timers
are used in scheduling the response. While a response timer is running 
at one node, if a response is received from another
node then the response timer is suppressed to reduce the number
of responses triggered. Consequently, the response time may be
delayed to allow for more suppression.

Two main performance evaluation criteria used in this case are
overhead of response messages and time to recover from packet loss.
Depending on the relative delays between group members and the
timer settings, the mechanism may exhibit different performance. In this
study, our method attempts to obtain scenarios of best case and worst
case performance according to the above criteria.

We are not aware of any related work that attempts to achieve this goal
systematically. However, we borrow from previous work on protocol verification and
test generation. 

The rest of the paper is organized as follows. 
Section~\ref{model} introduces the protocol and topology models.
Section~\ref{apply} outlines the main algorithm, and Section~\ref{timer} 
presents the model for TSM.
Sections~\ref{overhead} and ~\ref{response} present performance 
analyses for protocol overhead and response time, and Section~\ref{simulation} 
presents simulation results.
Related work is described in Section~\ref{related}. 
Issues and future work are discussed in 
Section~\ref{issues}. We present concluding remarks in Section~\ref{conclusion}.
Algorithmic details, mathematical models and example case studies are given 
in the appendices.

\section{The Model}
\label{model}

The model is a processable representation of the system under
study that enables automation of our method.
Our overall model consists of: A) the protocol model, B) the
topology model, and C) the fault model.

\subsection{The Protocol Model}
\label{fsm}

We represent the protocol by a finite state machine (FSM) and the
overall system by a global FSM (GFSM).

{\em I. FSM model:}
Every instance of the protocol, running on a single end-system,
is modeled by a deterministic FSM consisting of: (i) a set of
states, (ii) a set of stimuli causing state transitions, and
(iii) a state transition function (or table) describing the state
transition rules. A protocol running on an end-system $i$ is represented by the
machine
${\mathcal{M}}_{i} = ({\mathcal{S}}_{i},\tau_{i},\delta_{i})$, where
${\mathcal{S}}_{i}$ is a finite set of state symbols,
$\tau_{i}$ is the set of stimuli, and
$\delta_{i}$ is the state transition function
${\mathcal{S}}_{i} \times \tau_{i} \rightarrow {\mathcal{S}}_{i}$.

{\em II. Global FSM model:}
The global state is defined as the composition of individual
end-system states. The behavior of a system with $n$ end-systems
may be described by 
$\mathcal{M}_{\mathcal{G}} =
(\mathcal{S}_{\mathcal{G}},\tau_{\mathcal{G}},\delta_{\mathcal{G}})$,
where
$\mathcal{S}_{\mathcal{G}}$: ${\mathcal{S}}_{1} \times
{\mathcal{S}}_{2}
\times \dots \times {\mathcal{S}}_{n}$ is the global state space,
$\tau_{\mathcal{G}}$: $\overset{n}{\underset{i=1}{\bigcup}}
\tau_i$ is the set of stimuli, and
$\delta_{\mathcal{G}}$ is the global state transition function
$\mathcal{S}_{\mathcal{G}} \times \tau_{\mathcal{G}}
\rightarrow \mathcal{S}_{\mathcal{G}}$.

\subsection{The Topology Model}

\begin{figure}[th]
 \begin{center}
  \epsfig{file=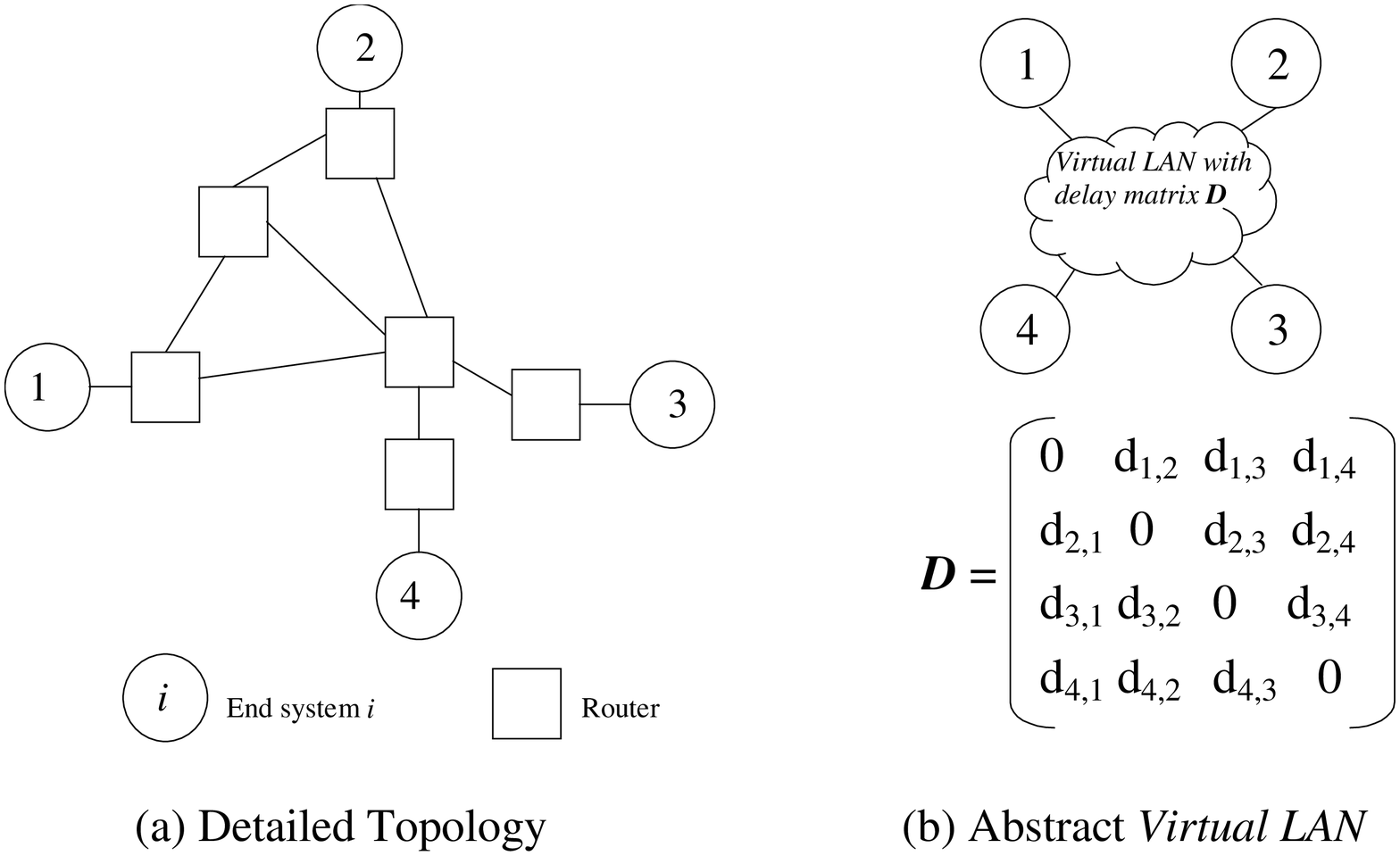,height=6.75cm,width=12cm,clip=,angle=0}
   \caption{Representing network details using {\em virtual LAN} and delay
matrix {\bf\em D}. ($d_{i,j}$ is delay from $i$ to 
$j$.)}\label{vlan_figure}
 \end{center}
\end{figure}

	The topology cannot be captured simply by one metric. Indeed, its
dynamics may be complex to model and sometimes intractable. We model the
topology at the network layer and we abstract the network using end-to-end
delays. We model the delays using the delay matrix and loss patterns using
the fault model. We use a {\em virtual LAN} (VLAN) model to represent the
underlying network topology and multicast distribution tree. The VLAN
captures delay semantics using a delay matrix $D$ (see
Figure~\ref{vlan_figure}), where $d_{i,j}$ is the delay from system $i$ to
system $j$\footnote{Throughout this documents, we use the term {\em
topology synthesis} to denote the assignment of delay values which 
constitute the entries of the $D$ matrix.}.
The VLAN model may seem as an over-simplification of the topology as it
abstracts the internal network connectivity and queues.  This, however,
renders our model tractable and is quite useful in obtaining
characteristics of boundary-point scenarios. We shall further investigate
the utility and accuracy of our model in Section~\ref{simulation} through
detailed packet level simulations of sophisticated timer mechanisms
over complex topologies.

\subsection{The Fault Model}

A {\em fault} is a low level (e.g., physical layer) anomalous
behavior that may affect the protocol under test. Faults may
include packet loss, system crashes, or routing loops. For
brevity, we only consider selective packet loss in this study. Selective
packet loss occurs when a multicast message is received
by some group members but not others. 
The selective loss of a message
prevents the transition that this message triggers at the intended recipient.

\section{Algorithm and Objectives}
\label{apply}

To apply our method, the designer specifies the protocol as a global FSM model.
In addition, the evaluation criteria, be it related to performance
or correctness, are given as input to the method.
In this paper we address performance criteria, correctness has been addressed in
previous studies~\cite{stress,fotg}.
The algorithm operates on the specified model and synthesizes a set of `test
scenarios'; protocol events and relations between topology delays and timer values,
that stress the protocol according to the evaluation criteria (e.g., exhibit maximum
overhead or delay).
In this section, we outline the algorithmic details of our method.
The algorithm is further discussed in Section~\ref{overhead} and
illustrated by a case study.
Algorithmic complexity issues are discussed in Section~\ref{issues}.

\subsection{Algorithm Outline}
\label{algorithm}

\begin{figure}[th]
 \begin{center}
\epsfig{file=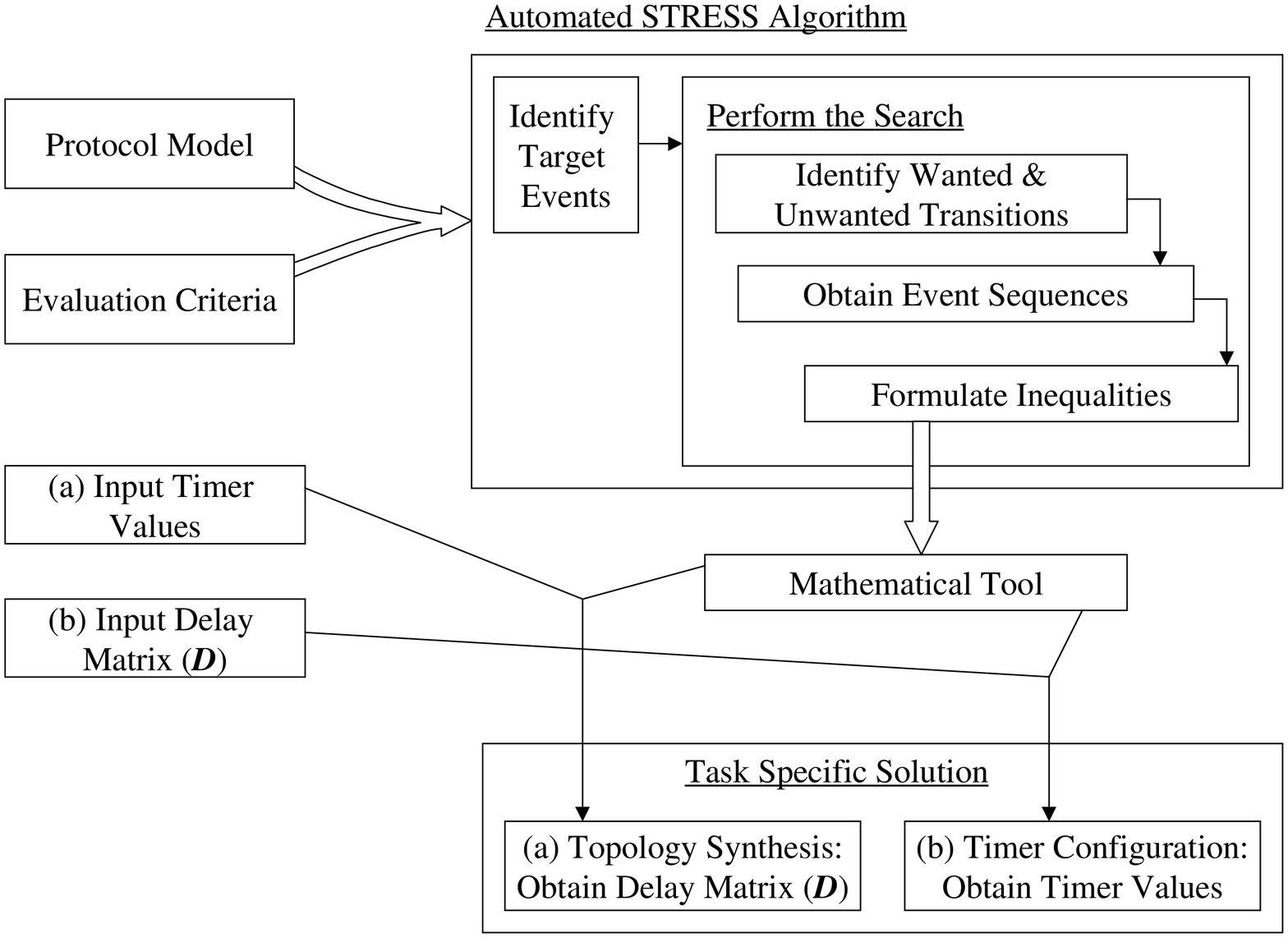,height=7cm,width=12cm,clip=,angle=0}
   \caption{STRESS framework for boundary-point 
scenario synthesis.}\label{block_diag}
 \end{center}
\end{figure}

Our algorithm is a variant of the fault-oriented test 
generation (FOTG) algorithm presented in~\cite{fotg}. It includes the 
topology synthesis, the backward search and the forward search stages.
Here we describe those aspects of our algorithm that deal
with timing and performance semantics.
The basic algorithm passes through three main steps (1) the target event 
identification, (2) the search, and (3) the task specific solution. The algorithm is
outlined in Figure~\ref{block_diag}.

\begin{enumerate}
\item {\bf The target event:}
The algorithm starts from a given
event, called the `target event'. The target event (e.g., sending a 
message) is identified 
by the designer based on the protocol evaluation criteria, e.g.,
overhead.

\item {\bf The search:} Three steps are taken in the search: 
(a)~identifying 
conditions, (b)~obtaining sequences, and (c)~formulating inequalities.

\begin{enumerate}
\item {\em Identifying conditions:}
The algorithm uses the protocol transition rules to identify 
transitions necessary to trigger the target event and those that prevent it, 
these transitions are called {\em wanted transitions} and {\em unwanted transitions},
respectively.

\item {\em Obtaining sequences:}
Once the above transitions are identified, the
algorithm uses backward and forward search
to build event sequences leading to these transitions and calculates 
the times of these events as follows. 
\begin{enumerate}
\item {\bf Backward search} is used to identify events preceding the 
wanted and unwanted transitions, and uses implication rules that operate
on the protocol's transition table. 
Section~\ref{implication_rules} describes the implication rules.
\item {\bf Forward search} is used to verify the backward search.
Every backward step must correspond to valid forward step(s). Branches 
leading to contradictions between forward and backward search are 
rejected. Forward search is also used to complete event sequences 
necessary to maintain system consistency\footnote{The role of forward 
search will be further illustrated in the response time analysis in 
Section~\ref{response}.}. 
\end{enumerate}
\item {\em Formulating inequalities:}
Based on the transitions and timed sequences obtained in the previous steps,
the algorithm formulates relations between timer values and network delays
that trigger the wanted transitions and avoid the unwanted transitions.
\end{enumerate}
\item {\bf Task specific solution:}
The output of the search is a set of event
sequences and inequalities that satisfy the evaluation criteria.
These inequalities are solved mathematically to find a
topology or timer configuration, depending on the task definition.
\end{enumerate}

\subsection{Task Definition}

We apply our method to two kinds of tasks:
\begin{enumerate}
\item {\bf Topology synthesis} is performed to identify the 
delays, $d_{i,j}$, in the
dealy matrix $D$ that produce the best or worst case
behavior, given the timer values\footnote{If the topology connectivity 
is also given, the task may also include obtaining {\em link} delays, not 
only end-to-end delays as in the $D$ matrix. For our discussion in this 
document we will assume that identifying the entries of the $D$ matrix is 
the task. Appendix B discusses the problem formulation 
to accommodate {\em link} delays.}. 
\item {\bf Timer configuration} is performed to 
obtain the timer values that cause
the best and worst case behavior, given the topology delay matrix $D$.
\end{enumerate}

\section{The Timer Suppression Mechanism (TSM)}
\label{timer}

In this section, we present a simple description of TSM,
then present its model, used thereafter in the analysis.
TSM involves a request, $q$, and one or
more responses, $p$. When a system, $Q$, detects the loss of a data
packet it sets a request timer and multicasts a request $q$.
When a system $i$ receives $q$ it sets a response timer (e.g., randomly), 
the expiration of which,
after duration $Exp_i$, triggers a response $p$. If the system
$i$ receives a response $p$ from another system $j$ while its timer
is running, it suppresses its own response.

\subsection{Performance Evaluation Criteria}

We use two performance criteria to evaluate TSM:
\begin{enumerate}
\item Overhead of response messages, where the worst case 
produces the maximum number of responses per data packet loss.
As an extreme case, this occurs when 
all potential responders respond
and no suppression takes place.
\item The response delay, where worst case scenario produces maximum loss recovery
time.
\end{enumerate}

\subsection{Timer Suppression Model}

Following is the TSM model used
in the analysis.

\subsubsection{Protocol states ($\mathcal{S}$)}

Following is the state symbol table for the TSM model.

\footnotesize

\hspace{-0.5cm}
\vspace{.1in}
\begin{tabular}{|ll|} \hline
State & Meaning \\ \hline \hline
$R$ & original state of the requester $Q$ \\
$R_T$ &  requester with the request timer set \\ 
$D$ & potential responder \\
$D_T$ & responder with the response timer set \\ \hline
\end{tabular}

\small

\subsubsection{Stimuli or events}

\begin{enumerate}
\item Sending/receiving messages: transmitting response 
($p_t$) and request ($q_t$), receiving
response ($p_r$) and request ($q_r$).
\item Timer events and other events:
the events of firing the request timer, $Req$, and
response timer, $Res$, and the event of
detecting packet loss, $L$.
\end{enumerate}

\subsubsection{Notation}

Following are the notations used in the transition table and the analysis 
thereafter.

\begin{itemize}
\item An event subscript
denotes the system initiating the event,
e.g.,
$p_{t_i}$ is response sent by system $i$, while the
subscript
$m$ denotes multicast reception, e.g., $p_{r_m}$ denotes scheduled reception
of a response by all members of the group if no loss occurs.
When system $i$ receives a message sent by system $j$, this is
denoted by the subscript $i,j$, e.g., $p_{r_{i,j}}$ denotes system
$i$ receiving response from system $j$.

\item The state subscript $T$
denotes the existence
of a timer,
and is used by the algorithm to apply the
{\em timer implication} to fire the timer event after the expiration 
period $Exp$.
\item A state transition has a $start$ state and an $end$ state and
is expressed in the form $startState \rightarrow endState$
(e.g. $D
\rightarrow D_T$). It implies the existence of a system in the
$startState$ (i.e., $D$) as a condition for the
transition to the $endState$ (i.e., $D_T$). 
\item {\em Effect} in the transition table may contain transition
and stimulus in the form  
($startState \rightarrow endState).stimulus$,
which indicates that the condition for
triggering $stimulus$ is the state transition.
An effect may contain several transitions
(e.g., $Trans1$, $Trans2$), which means that out of these
transitions only those with satisfied conditions will take effect.

\item To describe event sequences in the backward search we denote $a 
\Leftarrow b$, where $a$ and $b$ are global states, which means that $a$ 
succeeds $b$ in the event sequence, and that $b$ can be implied from $a$. 
Also, for forward search we use $a \Rightarrow b$, which means that $a$ 
precedes $b$ in the event sequence and that $b$ can be implied from $a$.

\end{itemize}

\subsubsection{Transition table}
\label{transition_table}

Following is the transition table for TSM.

\footnotesize

\hspace{-0.5cm}
\vspace{.1in}
\begin{tabular}{|llll|} \hline
Symbol & Event & Effect & Meaning \\ \hline \hline
$loss$ & $L$ & $(R \rightarrow R_T).q_t$ & Loss detection causes $q_t$ and setting of
request timer \\ \hline
$tx\_req$ & $q_{t}$ & $q_{r_m}$ 	& Transmission of $q$ causes multicast
reception of $q$ after network delay \\ \hline
$rcv\_req$ & $q_{r}$ & $D \rightarrow D_T$ & Reception of $q$ causes a system in $D$
state to set response timer \\ \hline
$res\_tmr$ & $Res$	& $(D_T \rightarrow D).p_{t}$ & Response timer expiration
causes transmission of $p$ and a change to $D$ state \\ \hline
$tx\_res$ & $p_{t}$ & $p_{r_m}$ 	& Transmission of $p$ causes multicast
reception of $p$ after network delay \\ \hline
$rcv\_res$ & $p_{r}$ & $R_T \rightarrow R$, $D_T \rightarrow D$ & Reception of $p$ by
a system with the timer set causes suppression \\ \hline
$req\_tmr$ & $Req$ & $q_t$ & Expiration of request timer causes 
re-transmission of $q$
\\ \hline
\end{tabular}

\small

The model contains a requester, $Q$, and several
potential responders (e.g., $i$ and $j$).\footnote{Since there
is only one requester, we simply use $q_t$ instead of $q_{t_Q}$,
$q_{r_i}$ instead of $q_{r_{i,Q}}$, $Req$ instead of $Req_Q$, $R$ instead of $R_Q$
and $R_T$ instead of $R_{T_Q}$.}
Initially, the requester, $Q$, exists in state $R$ and all potential responders
exist in state $D$. 
Let $t_0$ be the time at which $Q$ sends
the request, $q$.
The request sent by $Q$ is received by $i$ and $j$ at
times $d_{Q,i}$ and $d_{Q,j}$, respectively.
When the request, $q$, is sent, the requester transitions into state
$R_T$ by setting the request timer.
Upon receiving a request, a potential responder in state $D$
transitions into state $D_T$, by setting the response timer. 
The time at which an event occurs is given by $t(event)$, 
e.g., $q_{r_j}$ occurs at $t(q_{r_j})$.\footnote{The time of a
state is when the state was first created, so $t(D_{T_i})$ is
the time at which $i$ transited into state $D_T$.}

\subsubsection{Implication rules}
\label{implication_rules}

The backward search uses the following cause-effect implication rules:
	\begin{enumerate}
	\item Transmission/Reception ({\bf Tx\_Rcv}): By the
reception of a message, the algorithm implies the transmission of
that message --without loss-- sometime in the past (after
applying the network delays).
An example of this implication is $p_{r_{i,j}} \Leftarrow
p_{t_j}$, where $t(p_{r_{i,j}}) = t(p_{t_j}) + d_{j,i}$.
	\item Timer Expiration ({\bf Tmr\_Exp}): 
When a timer expires, the algorithm infers that it was set $Exp$
time units in the past, and that no event occurred during that
period to reset the timer.
An example of this implication is $Res_i.(D_i \leftarrow 
D_{T_i}) \Leftarrow D_{T_i}$, where $t(Res_i) = t(D_{T_i}) + Exp_i$,
and $Exp_i$ is the duration of the response timer
$Res_i$.\footnote{We use the notation $Event.Effect$ to 
represent a transition.}

	\item State Creation ({\bf St\_Cr}): 
To build a history of events leading to a certain state, we reverse
the transition rules and get to the $startState$ of the transitions leading to the
creation of the state in question. For example, $D_{T_i} \Leftarrow (D_{T_i}
\leftarrow D_i)$ means that for the system to be in state $D_{T_i}$ the system must
have existed in state $D_i$ and this is implied from the transition $(D_{T_i}
\leftarrow D_i)$.
	\end{enumerate}

In the following sections we use the above model to
synthesize worst and best case scenarios according to
protocol overhead and response time.

\section{Protocol Overhead Analysis}
\label{overhead}

	In this section, we conduct worst and best case
performance analyses for TSM with
respect to the number of responses triggered per packet loss.
Initially, we assume no loss of request or response messages until recovery, and
that the request timer is high enough that the
recovery will occur within one request round. 
The case of multiple request rounds is discussed in Appendix C.

\subsection{Worst-Case Analysis}
\label{worst_case}

Worst-case analysis aims to obtain
scenarios with maximum number of responses per data loss.
In this section we present the algorithm to obtain
inequalities that lead to worst-case scenarios. These inequalities
are a function of network delays and timer expiration values.

\subsubsection{Target event}
	Since the overhead in this case is measured as the number
of response messages, the designer identifies the event of
triggering a response, $p_t$, as the target event, and the goal is to 
maximize the number of response messages. 

\subsubsection{The search}

As previously described in Section~\ref{algorithm},
the main steps of the search algorithm are to: 
(1) identify the wanted and unwanted transitions,
(2) obtain sequences leading to the above transitions,
and calculating the times for these sequences, and
(3) formulate the inequalities that achieve the time constraints 
required to invoke wanted transitions and avoid unwanted transitions.

\begin{figure}[th]
 \begin{center}
  \epsfig{file=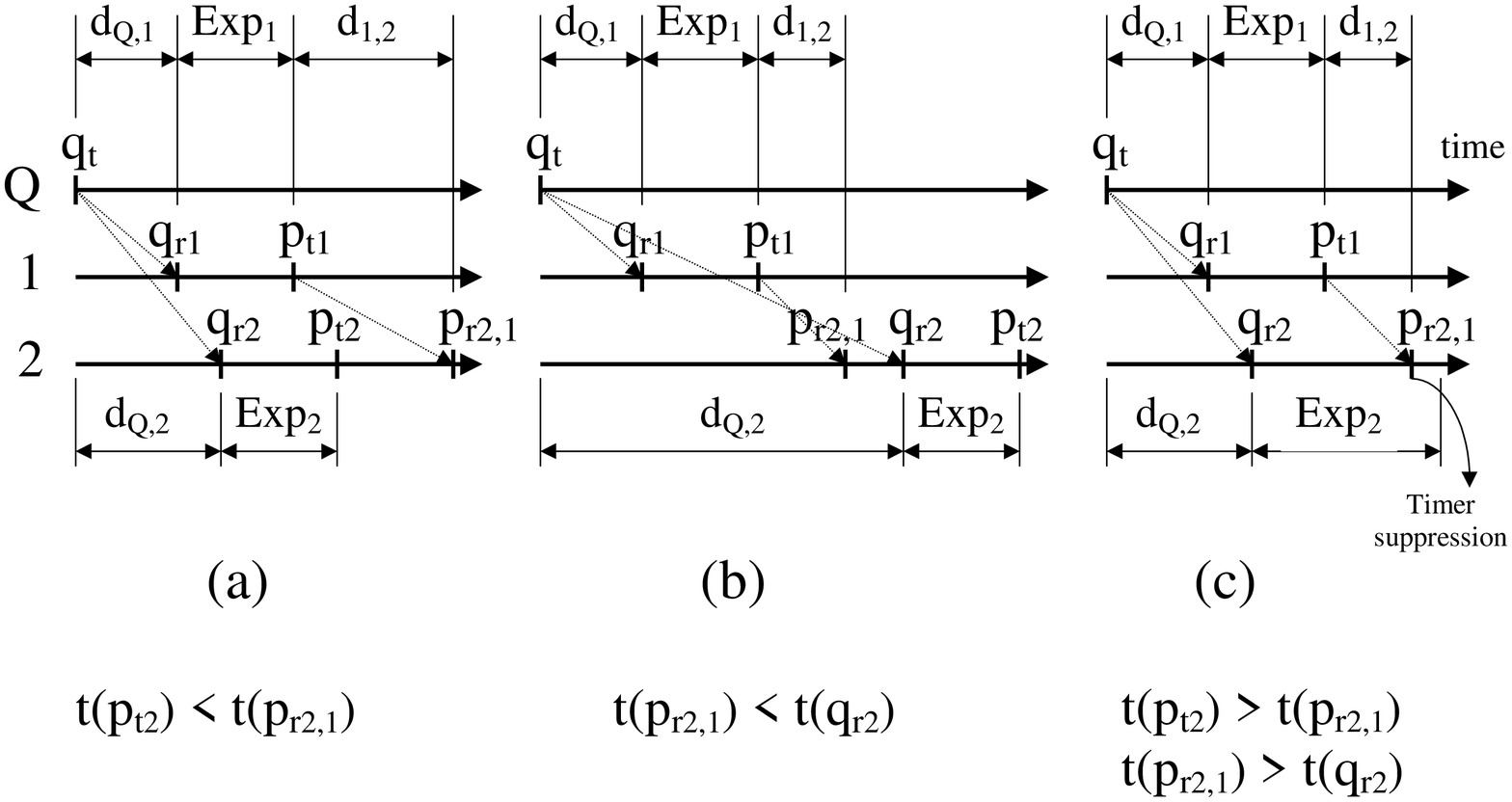,height=6cm,width=9cm,clip=,angle=0}
   \caption{Event sequencing: sequences (a) \& (b) do not lead 
to suppression, whereas (c) leads to suppression.}\label{time_fig}
 \end{center}
\end{figure}

\begin{itemize}
\item {\bf Identifying conditions:}
The algorithm searches for the transitions necessary to trigger the target 
event, and their conditions, recursively.
These are called {\em wanted transitions} and {\em wanted conditions},
respectively.
The algorithm also searches for transitions that nullify the 
target event or invalidate any of its conditions. These are called 
{\em unwanted transitions}.

In our case the target event is the transmission of a response (i.e., $p_t$). 
From the transition table described in Section~\ref{transition_table},
the algorithm identifies transition {\em res\_tmr}, or $Res.(D_T \rightarrow
D).p_{t}$, as a {\em wanted transition} and its condition $D_T$ as a
{\em wanted condition}. Transition {\em rcv\_req}, or $q_{r}.(D 
\rightarrow D_T)$, is also identified 
as a {\em wanted transition} since it is necessary to create $D_T$.
The {\em unwanted transition} is identified as transition {\em rcv\_res}, 
or $p_{r}.(D_T \rightarrow D)$,
since it alters the $D_T$ state without invoking $p_t$.

\item {\bf Obtaining sequences:}
Using backward search, the algorithm obtains sequences and calculates 
time values for the following transitions: (1)~wanted transition, {\em 
res\_tmr}, (2)~wanted transition {\em rcv\_req}, and (3)~unwanted 
transition {\em rcv\_res}, as follows:

\begin{enumerate}
\item To obtain the sequence of events for transition {\em
res\_tmr}, the 
algorithm applies implication rules (see Section~\ref{implication_rules}) Tmr\_Exp,
St\_Cr, Tx\_Rcv in that 
order, and we get $res\_tmr_i \Leftarrow rcv\_req_i \Leftarrow tx\_req$, or 

\begin{center}
	$Res_i.(D_i \leftarrow D_{T_i}).p_{t_i}
\Leftarrow q_{r_i}.(D_{T_i} \leftarrow D_i) 
\Leftarrow q_{t}$.
\end{center}

Hence the calculated time for $t(p_{t_i})$ becomes 

\[ t(p_{t_i}) = t_0 + d_{Q,i} + Exp_i, \]

where $t_0$ is the time at which $q_{t}$ occurs.

\item To obtain the sequence of events for transition {\em rcv\_req}
the algorithm applies implication rule Tx\_Rcv, and we get $rcv\_req_i \Leftarrow
tx\_req$, or 

\begin{center}
$q_{r_i}.(D_{T_i} \leftarrow D_i) \Leftarrow
q_{t}$.
\end{center}

Hence the calculated time for $t(q_{r_i})$ becomes

\[ t(q_{r_i}) = t_0 + d_{Q,i}. \]

\item To obtain sequence of events for transition {\em rcv\_res} for
systems $i$ and $j$ the algorithm applies implication rules 
Tx\_Rcv,Tmr\_Exp, St\_Cr, Tx\_Rcv in that order, and we get $rcv\_res_i \Leftarrow
res\_tmr_j \Leftarrow rcv\_req_j \Leftarrow tx\_req$, or

\begin{center}
$p_{r_{i,j}}.(D_i \leftarrow D_{T_i}) \Leftarrow
Res_j.(D_j \leftarrow D_{T_j}).p_{t_j} \Leftarrow
q_{r_j}.(D_{T_j} \leftarrow D_j) \Leftarrow
q_{t}$.
\end{center}

Hence the calculated time for $t(p_{r_{i,j}})$ becomes

\[ t(p_{r_{i,j}}) = t_0 + d_{Q,j} + Exp_j + d_{j,i}. \] 

\end{enumerate}

\item {\bf Formulating Inequalities:}
Based on the above wanted and unwanted transitions
the algorithm forms constraints and conditions to aoivd the unwanted transition, {\em
rcv\_res}, while invoking the wanted transition,
{\em res\_tmr}, to transit out of $D_T$. To achieve this, the algorithm 
automatically derives the following inequality (see Appendix A
for more details):

\begin{equation}
	t(p_{t_i}) < t(p_{r_{i,j}}).
\end{equation} 

	Substituting expressions for $t(p_{t_i})$ and
$t(p_{r_{i,j}})$ previously derived, we get:

\begin{center}
	$d_{Q,i} + Exp_i < d_{Q,j} + Exp_j + d_{j,i}$.
\end{center}

Alternatively, we can avoid the unwanted transition {\em rcv\_res} if the system did
not exist in $D_T$ when the response is received.
Hence, the algorithm automatically derives the following inequality (see Appendix A
for more details):

\begin{equation}
	t(p_{r_{i,j}}) < t(q_{r_i}).
\end{equation}

Again, substituting expressions derived above, we get:

\begin{center}
	$d_{Q,i} > d_{Q,j} + Exp_j + d_{j,i}$.
\end{center}	

Note that equations (1) and (2) are general for any number of
responders, where $i$ and $j$ are any two responders in the
system.
Figure~\ref{time_fig} (a) and (b) show equations (1) and (2), respectively. 

\end{itemize}

\subsubsection{Task specific solutions}

\begin{itemize}
\item {\bf Topology synthesis:}
Given the timer expiration values or ranges we want to
find a feasible solution for the worst-case delays. A feasible solution in
this context means assigning positive values to the delays $d_{i,j} \forall i, j$.

In equation (1) above, if we take $d_{Q,i} = d_{Q,j}$\footnote{The number of
inequalities ($n^2$, where $n$ is the number of responders) is less then the
number of the unknowns $d_{i,j}$ ($n^2 - n$), hence there are 
multiple solutions. We can obtain a solution by assigning values 
to $n$ unknowns (e.g., $d_{Q,i}$) and solving for the others.}, we get:

\begin{center}
$Exp_i - Exp_j < d_{j,i}$.
\end{center}

These inequalities put a lower limit on the delays $d_{j,i}$,
hence, we can always find a positive $d_{j,i}$ to satisfy the
inequalities.
Note that, the delays used in the delay matrix reflect delays over
the multicast distribution tree. In general, these delays are
affected by several factors including the multicast and unicast routing
protocols, tree type and dynamics, propagation, transmission and queuing
delays.
	One simple topology that reflects the delays of the
delay matrix is a completely connected network where the underlying 
multicast distribution tree coincides with the unicast routing.
There may also exist many other complex topologies that satisfy the delay matrix
$D$\footnote{Mapping from the delay matrix $D$ into complex topologies is not covered
in this document.}.

\item {\bf Timer configuration:}
Given the delay values, ranges or bounds, we want to obtain timer 
expiration values that produce worst-case behavior.
We obtain a range for the relative timer 
settings (i.e., $Exp_i - Exp_j$) using equation (1) above. 

\end{itemize}

The solution for the system of inequalities given by (1) and (2) above can be solved
in
the general case using linear programming (LP) techniques (see Appendix B for more
details). Section~\ref{simulation} uses the above solutions to synthesize simulation
scenarios.

Note, however, that it may not be feasible to satisfy all these
constraints, due to upper bounds on the delays for example. 
In this case the problem becomes one of maximization, where
the worst-case scenario is one that triggers maximum number
of responses per packet loss. This problem is discussed in 
Appendix B. 

\subsection{Best-Case Analysis}
\label{best_case}

Best case overhead analysis constructs constraints that lead to
maximum suppression, i.e., minimum number of responses.
The following conditions are formulated using steps similar to
those given in the worst-case analysis:

\begin{equation}
	t(p_{t_i}) > t(p_{r_{i,j}}),
\end{equation}

and

\begin{equation}
	t(p_{r_{i,j}}) > t(q_{r_i}).
\end{equation}

These are complementary conditions to those given in the worst
case analysis. Figure~\ref{time_fig} (c) shows equations (3) and (4). 
Refer to the Appendix A 
for more details on the inequality derivation.

This concludes our description of the algorithmic details to construct worst and
best-case delay-timer relations for overhead of response messages. Solutions to these
relations represent delay and timer settings for stress scenarios that are used later
on for simulations. 

\section{Response Time Analysis}
\label{response}

	In this section, we conduct the performance analysis 
with respect to response time, i.e., the time for the requester to
recover from the packet loss. 
The algorithm obtains possible sequences leading to the target
event and calculates the response time for each sequence.
To synthesize the worst case scenario that
maximizes the response time, for example, the sequence with maximum
time is chosen.

To systematically approach this problem we consider the following three cases: (1) The
case of {\em no} loss to the response message. This case leads to single round of
request-response messages. Without loss of response messages this problem becomes one
of maximizing the round trip delay between the requester and the first responder.
(2) The case of single selective\footnote{In selective loss the response may be
received by some systems but not others.} loss of the response message. This case
may lead to two rounds of request-response messages. We analyze this case in the
first part of this section.
(3) The case of multiple selective losses of the response messages. This case
may lead to more than two rounds of request-response messages, and is discussed at
the end of this section.

We now consider the case of {\bf single selective loss} of the 
response message during the recovery phase. For selective losses, transition rules are
applied to only those systems that receive the message.

\subsection{Target Event}

The response time is the time taken by the mechanism to recover
from the packet loss, i.e., until the requester receives the
response $p$ and resets its request timer by transitioning out of
the $R_T$ state. In other words,
the response interval is $t(p_{r_Q}) - t(q_{t}) = t(p_{r_Q}) - t_0$.
The designer identifies $t(p_{r_Q})$ as the target time, hence,
$p_{r_Q}$ is the target event.

\subsection{The Search}

We present in detail the case of single
responder, then discuss the multiple responders case.

\begin{itemize}
\item {\bf Backward search:}
As shown in Figure~\ref{response_diag} (a), the backward search starts 
from $p_{r_Q}$
and is performed over the transition table 
(see Section~\ref{transition_table}) using
the implication rules in Section~\ref{implication_rules},
yielding $rcv\_res_Q \Leftarrow res\_tmr_j \Leftarrow rcv\_req_j$, or~\footnote{The
GFSM may 
be represented by composition of individual states 
(e.g., $State_1.State_2$ or $transition_1.State_2$).}:

\begin{center}
$D_j.p_{r_Q}.(R \leftarrow R_T) \Leftarrow p_{t_j}.(D_j
\leftarrow D_{T_j}).Res_j.R_T \Leftarrow q_{r_j}.(D_{T_j}
\leftarrow D_j).R_T$
\end{center}

At which point the algorithm reaches a branching point, where two
possible preceding states could cause $q_{r_j}$:
\begin{itemize}
\item The first is transition {\em loss}, or $D_j.q_{t}.(R_T \leftarrow 
R)$, and since the initial state $R$ is reached,
the backward search ends for this branch. 
\item The second is transition {\em req\_tmr}, or $D_j.Req.q_{t}.R_T$.
Note that $Req$ indicates the need for a transition to $R_T$, i.e., 
($R_T \leftarrow R$),  
and the search for this last state yields the intial (data packet loss) state
$loss$: $D_j.q_{t}.(R_T \leftarrow R)$. However, $q_t$ is message transmission, which
implies that the message must be received (or lost). Hence, there are gaps in the
event sequence (indicated by the $dots$ in
Figure~\ref{response_diag} (a)) that are filled through forward
search (in Figure~\ref{response_diag} (b)).
\end{itemize}

\begin{figure}[th]
 \begin{center}
\epsfig{file=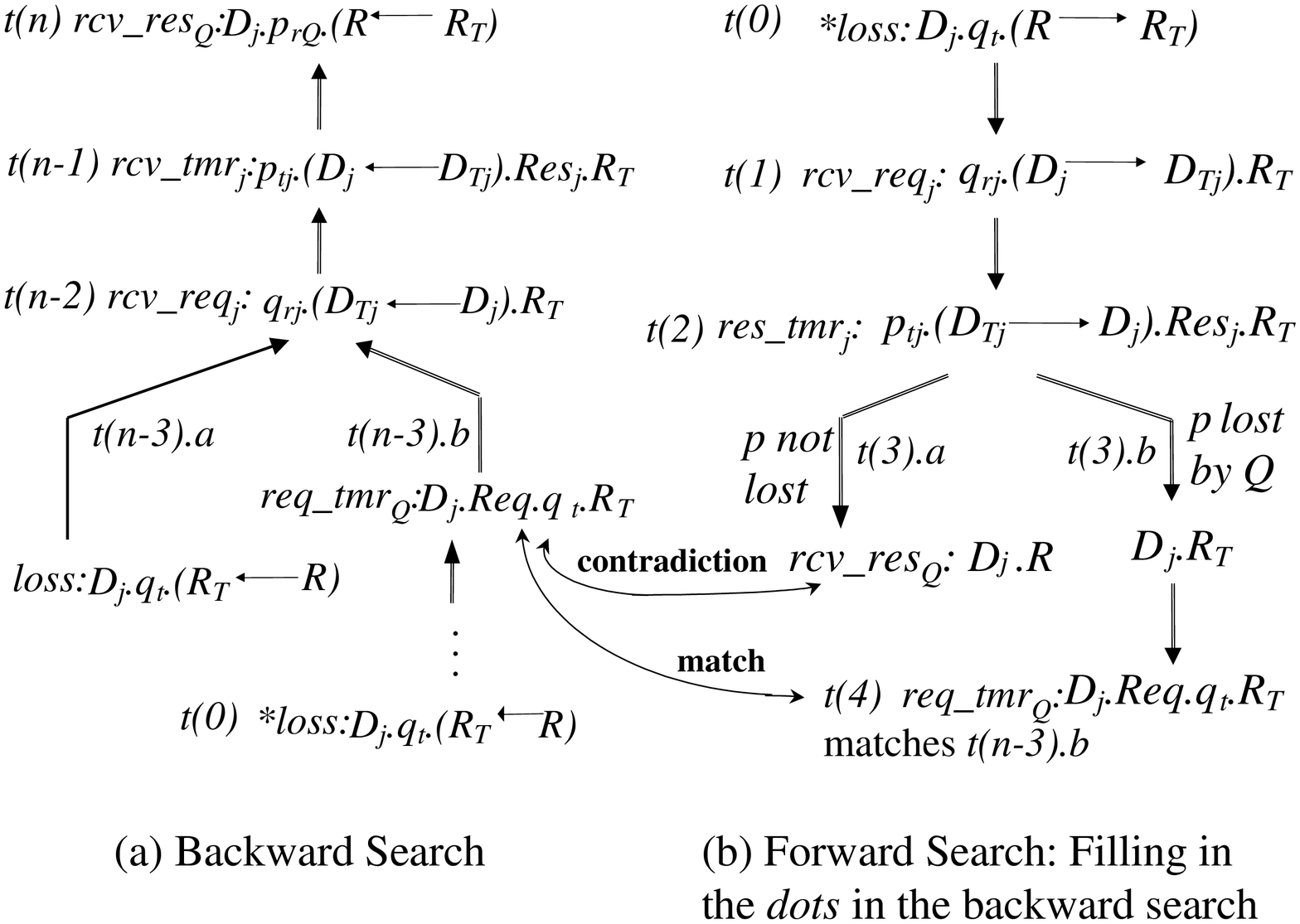,height=7cm,width=12cm,clip=,angle=0}
   \caption{Backward and forward search for response time
analysis: (a) Backward search starts from the loss recovery
state (at time $t(n)$) when the requester receives a response ($rcv\_res_Q$), and ends
at the
initial (packet $loss$) state at time $t(0)$. Part of the event sequence is
incomplete (denoted by $dots$). (b) Forward search starts (at time $t(0)$) from
the last state reached by the backward search after the $dots$
(denoted by $*loss$). It attempts to fill the gap in the sequence while checking 
for consistency, and finds a match.}\label{response_diag}
 \end{center}
\end{figure}

\item {\bf Forward search:}
The algorithm performs a forward search 
and checks for consistency of the GFSM. 
The forward search step may lead to contradiction with the
original backward search, causing rejection of that branch as a
feasible sequence. For example, as shown in Figure~\ref{response_diag} 
(b), one possible forward sequence from the 
initial state gives $loss \Rightarrow tx\_req \Rightarrow rcv\_req_j \Rightarrow
res\_tmr_j \Rightarrow tx\_res_j$, or:

\begin{center}
$D_j.q_{t}.(R \rightarrow R_T) \Rightarrow
q_{r_j}.(D_j \rightarrow D_{T_j}).R_T \Rightarrow
p_{t_j}.(D_{T_j} \rightarrow D_j).Res_j.R_T$
\end{center}


The algorithm then searches two possible next states:
\begin{itemize}
\item
If $p_{t_j}$ is not lost, and hence causes $p_{r_Q}$, then the next 
state is $D_j.R$. But
the original backward search started from 
$D_j.q_{t}.Req.R_T$
which cannot be reached from $D_j.R$. Hence, we get contradiction and 
the algorithm rejects this sequence.

\item If the response $p$ is lost by $Q$,
we get $D_j.R_T$ that leads to
$D_j.Req.q_{t}.R_T$.
The algorithm identifies this as a feasible sequence.
\end{itemize}
Calculating the time for each feasible sequence, the algorithm
identifies the latter sequence as one of maximum response time. 
\end{itemize}

For {\bf multiple responders}, the algorithm automatically explores
the different possible selective loss patterns of the response
message. The search identifies the sequence with maximum response as one
in which only one responder triggers a response that is selectively lost
by the requester.
To construct such a sequence, the algorithm creates conditions and 
inequalities similar to those formulated for the
best-case overhead analysis with respect to number of responses (see 
Section~\ref{best_case}). 

Effectively, the sequence obtained above occurs when the response is lost by the 
requester, which triggers another request.
Intuitively, the response delay is increased with multiple request rounds. 
The case of {\bf multiple selective loss} of the response messages may trigger
multiple (more than two) request rounds. Practically, the number of request rounds is
bounded by the protocol implementation, which imposes an upper bound on the number of
requests sent per packet loss. This, in turn, imposes an upper bound on the worst-case 
response time. This bound can be easily integrated into the search to end 
the search when the maximum number of allowed request rounds is reached\footnote{
The theoretical, trivial, worst-case response time is an infinite number of request
rounds. The goal of this analysis, however, is to provide a scenario in which
response time is maximized. It was a finding of our algorithm that if multiple rounds
are forced then the response time increases. It was also part our algorithm to
formulate conditions under which multiple response rounds are forced.}.

After conducting the above analyses, we have applied our method to generate worst-case
overhead scenarios for topology synthesis and timer configuration tasks using
determinsitic and adaptive timers (see Appendix D). We also applied it to response
time analysis and to best-case analyses. In the next section we show network
simulations using our generated worst-case overhead scenarios. 

\section{Simulations Using Systematic Scenarios}
\label{simulation}

To evaluate the utility and accuracy of our method, we have conducted a 
set of detailed simulations for
the Scalable Reliable Multicast (SRM)~\cite{SRM} based on our worst-case
scenario synthesis results for the timer-suppression mechanism. 
We tied our method to the network simulator (NS)~\cite{ns}.  The output
of our method, in the form of inequalities (see Section~\ref{overhead}), is solved
using a mathematical package (LINDO).  The solution, in terms of a delay
matrix, is then used to generate the simulation topologies for NS automatically.

\begin{figure}[th]
 \begin{center}
\epsfig{file=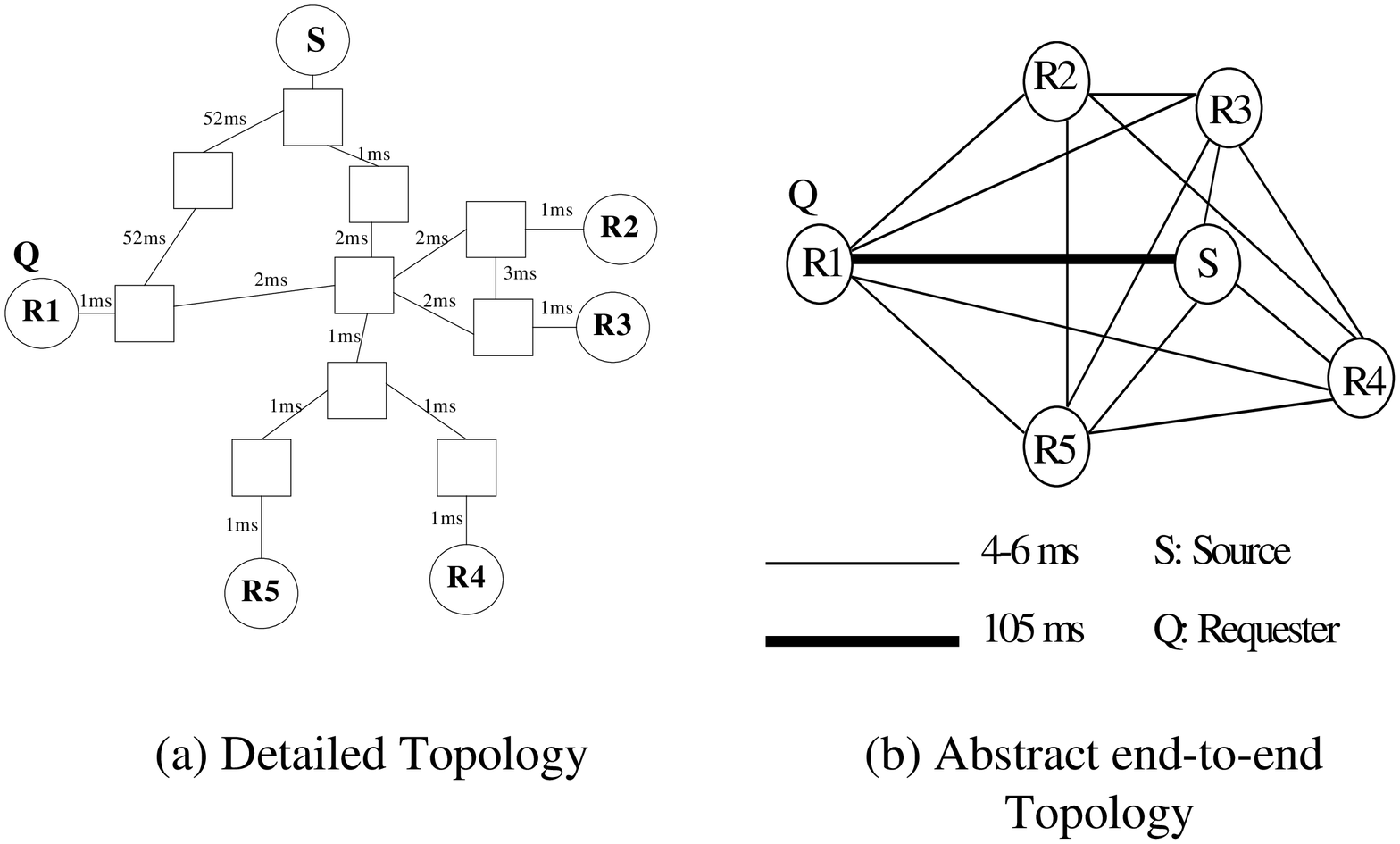,height=7.75cm,width=14cm,clip=,angle=0}
   \caption{Example {\em stress} topology used for the simulation.}\label{stress_topo}
 \end{center}
\end{figure}

For our simulations we measured the number of responses triggered for each data packet
loss. 
We have conducted two sets of simulations, each using two sets of topologies.
The simulated topologies included topologies with up to 200 receivers.  
The first set of
topologies was generated according to the overhead analysis presented in this paper. 
We call this set of topologies the {\em stress} topologies. Example {\em 
stress} topology is shown in Figure~\ref{stress_topo} (a), and its 
corresponding fully-connected topology is shown in 
Figure~\ref{stress_topo} 
(b). Both topologies satisfy the delay matrix, $D$, produced by 
stress\footnote{One may perceive the fully-connected graph as an 
abstraction of more complex topologies that satisfy the same delay 
matrix, $D$. Mapping of delay matrix into complex topologies is out of 
scope of this document.}. The second set of topologies was generated by 
the GT-ITM 
topology generator~\cite{gt_itm}, generating random and transit stub
topologies\footnote{This topology generator is probably representative of 
a standard tool for topology generation used in networking research. Using 
GT-ITM we have covered
most topologies used in several SRM studies~\cite{kannan}~\cite{poly}.}. 
We call
this set of topologies the {\em random} topologies\footnote{We faced difficulties
when choosing the lossy link for the {\em random} topologies
in order to maximize the number of responses.  This is an example
of the difficulties networking researchers face when trying to stress
networking protocols in an ad-hoc way.}.

\begin{figure}[t]
 \begin{center}
  \epsfig{file=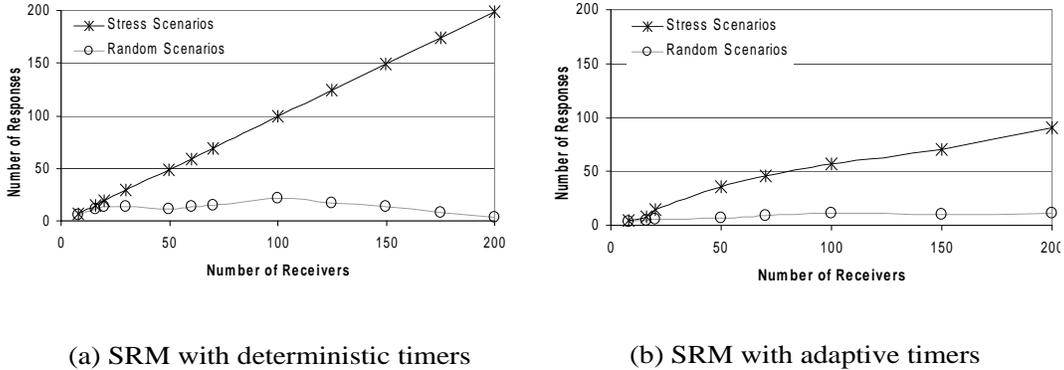,height=10cm,width=15cm,clip=,angle=0}
\vspace{-3.4cm}
   \caption{Simulation results for deterministic and adaptive timers over {\em stress}
and {\em random} topologies.}\label{sim_res1}
 \end{center}
\end{figure}

The first set of simulations was conducted for the SRM deterministic
timers\footnote{SRM response timer values are selected randomly from the interval
[$D_1.d_r$,$(D_1 + D_2).d_r$], where $d_r$ is the estimated distance to the requester,
and $D_1$, $D_2$ depend on the timer type. For deterministic timers $D_2=0$ and
$D_1=1$.}.
The results of the simulation are 
shown in
Figure~\ref{sim_res1} (a). The number of responses triggered for all the $stress$
topologies was $n-1$, where $n$ is the number of receivers
(i.e., no suppression occurred).  For the $random$ topologies, with up to 
200 receivers, the
number of responses triggered was less than 20 responses in the worst 
case.

Using the same two sets of topologies, the second set of simulations was
conducted for the SRM adaptive timers\footnote{Adaptive timers adjust their interval
based on the number of duplicate responses received and the estimated distance to the
requester.}.  The results are given in Figure~\ref{sim_res1} (b).  For the
$stress$ topologies almost 50\% of the receivers
triggered responses.  Whereas $random$ topologies
simulation generated almost 10 responses in the worst case, for 
topologies with 100-200 receivers.

These simulations illustrate how our method may be used
to generate consistent worst-case scenarios in a scalable fashion.  It
is interesting to notice that worst-case topologies generated for simple
deterministic timers also experienced substantial overhead (perhaps not the worst,
though) for more complicated timers (such as the adaptive timers).
It is also obvious from the simulations that {\em stress} scenarios are
more consistent than the other scenarios when used to compare
different mechanisms, in this case deterministic and adaptive timers; the performance
gain for adaptive timers is very clear under {\em stress} scenarios.

So, in addition to experiencing the worst-case behavior of a mechanism,
our stress methodology may be used to compare protocols in the above fashion and to
aid in investigating design trade-offs. It is a useful tool for generating 
meaningful
simulation scenarios that we believe should be considered in performance evaluation of
protocols in addition to the average case performance and random simulations.
We plan to apply our method to test a wider range of protocols through 
simulation\footnote{We have conducted other case studies using our 
STRESS method on multicast routing (PIM-DM~\cite{fotg}~\cite{stress_ic3n}, 
PIM-SM~\cite{stress}), 
MARS, Mobile-IP~\cite{mars}, and multicast-congestion 
control~\cite{stress_pgmcc}~\cite{sim_pgmcc}. We 
are currently
investigating ad hoc network protocols (e.g., MAC layer and ad hoc routing).}.


\section{Related Work}
\label{related}

Related work falls mainly in the areas of protocol verification, VLSI test 
generation and network simulation.

There is a large body of literature dealing with verification of
protocols. Verification systems typically address well-defined properties 
--such as {\em safety}, {\em liveness}, and {\em  
responsiveness}~\cite{proto_design}-- and aim to detect violations of 
these properties.
In general, the two main approaches for protocol verification are
theorem proving and reachability analysis~\cite{formal_survey1}.
Theorem proving systems define a set of axioms and relations to prove 
properties, and include {\em model-based} and {\em 
logic-based} formalisms~\cite{nqthm,z}.
These systems are useful in many applications.
However, these systems tend to abstract out some network dynamics that 
we study (e.g., selective packet loss).
Moreover, they do not synthesize network topologies and do not address 
performance issues per se.

Reachability analysis algorithms~\cite{reachability}, on
the other hand, try to inspect reachable protocol states, and
suffer from the `state space explosion' problem.
To circumvent this problem, state reduction
techniques could be used~\cite{partial_reachability2}.
These algorithms, however, do not synthesize network topologies.
Reduced reachability analysis has been used in the  verification of cache
coherence protocols~\cite{cache_coherence}, using a global FSM model.
We adopt a similar FSM model and extend it for our approach in this study.
However, our approach differs in that we address end-to-end protocols, 
that encompass rich timing, delay, and loss semantics, and we address 
performance issues (such as overhead or response delays).

There is a good number of publications dealing with conformance 
testing~\cite{Yannakakis}~\cite{conformance}~\cite{conformance1}~\cite{conformance2}.
However, conformance testing verifies that an implementation (as a black box) adheres
to a given specification of the protocol by constructing input/output sequences. 
Conformance testing is useful during the
implementation testing phase --which we do not address in this paper--
but does not address performance issues nor topology synthesis for design 
testing. 
By contrast, our method synthesizes test scenarios for protocol design, 
according to evaluation criteria. 

Automatic test generation techniques have been used in several fields.
VLSI chip testing~\cite{testability} uses test vector generation to detect
target faults. Test vectors may be generated based on circuit and 
fault models, using the fault-oriented technique, that utilizes {\em 
implication} techniques. These techniques were adopted in~\cite{fotg}
to develop fault-oriented test generation (FOTG) for multicast routing.
In~\cite{fotg}, FOTG was used to study correctness of a multicast routing
protocol on a LAN. We extend FOTG to study performance of end-to-end 
multicast mechanisms. We introduce the concept of a virtual LAN to 
represent the underlying network, integrate timing and delay semantics 
into our model and use performance criteria to drive our synthesis algorithm.

In~\cite{stress}, a simulation-based stress testing framework
based on heuristics was proposed. However, that method does not provide
automatic topology generation, nor does it address performance issues. 
The VINT~\cite{vint} tools provide a framework for Internet protocols 
simulation. Based on the network simulator (NS)~\cite{ns} and the network
animator (NAM)~\cite{nam}, VINT
provides a library of protocols and a set of validation test suites. 
However, it does not provide a generic tool for generating these tests 
automatically.
Work in this paper is complementary to such studies, and may be
integrated with network simulation tools similar to our work in 
Section~\ref{simulation}.

\section{Issues and Future Work}
\label{issues}

In this paper we have presented our first endeavor to automate
the test synthesis as applies to boundary-point performance evaluation of
multicast timer suppression protocols. Our case studies were by no means exhaustive.
However, they gave us insights into the research issues involved. Particularly, in
this section we shall discuss issues of algorithmic complexity. In addition, we
present our future plans to explore several potential extensions and applications of
our method.

\begin{itemize}

\item {\bf Algorithmic complexity}

One goal of our case studies is to understand and evaluate the
computational complexity of our method and algorithms. Our main algorithm uses a mix
of backward and forward search techniques. The algorithm starts from {\em target
events} and uses implicit backward search and branch and bound techniques to
synthesize the required scenario sequences.
Complexity of such algorithm depends on the finite state machine (FSM), the
state transition rules, and the target events from which the algorithm starts. Hence,
it is hard to quantify, in general terms, the complexity of our algorithm.
Nonetheless, we shall comment on the nature of the method and the algorithm
qualitatively based on our case studies. We note the following:
(a)~Our algorithms use branch and bound techniques and utilize implicit backward
search starting from a {\em target} event (vs. explicit forward reachability analysis
starting from initial states). 
Branch and bound techniques are, generally, hard to quantify in terms of (worst-case
or average case) complexity in abstract terms.
Although the worst-case for branch and bound could be
exponential, through our experiments we found that, on average, the target-based
approach has far less complexity than forward search.
In many cases the branch bounds immediately (e.g., due to contradiction if the
sequence is not feasible). 
For all our STRESS case studies, we have found our search algorithms to be quite
manageable.
(b)~Scenarios synthesized using the STRESS method usually are simple
and include relatively small topologies. Thus, they often experience low computational
complexity. It is our observation, in all our case studies thus far, that erroneous
and worst-case protocol behaviors may be invoked using relatively simple (yet
carefully synthesized) scenarios. Also, it was often the case that these simple
scenarios were extensible to larger and more complex scenarios using simple
heuristics. In Section~\ref{simulation}, we have demonstrated
how the simple scenarios generated by STRESS, with only a few receivers, could be
scaled up to include hundreds of receivers. Accuracy of
such extrapolation was validated through detailed simulations.

\item {\bf Automated generation of simulation test suites}

Simulation is a valuable tool for designing and evaluating
network protocols. Researchers usually use their
insight and expertise to develop simulation inputs and test
suites.
Our method may be used to assist in automating the process
of choosing simulation inputs and scenarios.
	The inputs to the simulation may include the topology,
host events (such as traffic models), network dynamics
(such as link failures or packet loss) and membership distribution and dynamics.
	Our future work includes implementing a more complete tool to
automate our method (including search algorithms and modeling semantics) and 
tie it to a network simulator to be applied to a wider range of 
multicast protocols.

\item {\bf Validating protocol building blocks}

	The design of new protocols and applications often
borrows from existing protocols or mechanisms. Hence, there is
a good chance of re-using established mechanisms, as appropriate,
in the protocol design process. Identifying, verifying and understanding 
building blocks for such mechanisms is necessary to increase
their re-usability. 
Our method may be used as a tool to improve that understanding in
a systematic and automatic manner.
	Ultimately, one may envision that a library of these building blocks
will be available, from which protocols (or parts thereof) will
be readily composable and verifiable using CAD tools; similar to
the way circuit and chip design is carried out today using VLSI
design tools.
In this work and earlier works~\cite{fotg}~\cite{stress}, some mechanistic
building blocks for multicast protocols were identified, namely, the
timer-suppression mechanism and the Join/Prune mechanism (for multicast routing).
More work is needed to identify more building blocks to cover a wider range of
protocols and mechanisms.

\item {\bf Generalization to performance bound analysis}

An approach similar to the one we have taken in this paper may be
based on performance bounds, instead of worst or best
case analyses. 
We call such approach `condition-oriented test generation'.

For example, a target event may be defined as
`the response time exceeding certain delay bounds' (either 
absolute or parametrized bounds).
	If such a scenario is not feasible, that indicates that
the protocol gives absolute guarantees (under the assumptions of
the study). This may be used to design and analyze quality-of-service
or real-time protocols, for example.

\item {\bf Applicability to other problem domains}

So far, our method has been applied mainly to case studies on multicast 
protocols in the context of the Internet.

Other problem and application domains may introduce
new mechanistic semantics or assumptions about the system or
environment. One example of such  domains includes sensor
networks. These networks, similar to 
ad-hoc networks, assume dynamic topologies, lossy channels, and
deal with stringent power constraints, which differentiates
their protocols from Internet protocols~\cite{sensornets}.

Possible research directions in this respect include:
\begin{itemize}
\item Extending the topology representation or model to capture
dynamics, where delays vary with time.
\item Defining new evaluation criteria that apply to the specific
problem domain, such as power usage.
\item Investigating the algorithms and search techniques that best
fit the new model or evaluation criteria.
\end{itemize}

\end{itemize}

\section{Conclusion}
\label{conclusion}

	We have presented a methodology for scenario synthesis for
boundary-point performance evaluation of multicast protocols.  In this
paper we applied our method to worst and best-case evaluation of the timer
suppression mechanism; a common building block for various multicast
protocols. We introduced a virtual LAN model to represent the underlying 
network topology and an extended global FSM model to represent the 
protocol mechanism. We adopted the fault-oriented test generation 
algorithm for search, and extended it to capture timing/delay semantics 
and performance issues for end-to-end multicast protocols.

	Two performance criteria were used for evaluation of the worst 
and best case scenarios; the number of responses per packet loss, and the 
response delay. Simulation results illustrate how our method can be used
in a scalable fashion to test and compare reliable multicast protocols.

	We do not claim to have a generalized algorithm that applies to
any arbitrary protocol. However, we hope that similar approaches may be
used to identify and analyze other protocol building blocks. We believe
that such systematic analysis tools will be essential in designing and
testing protocols of the future.

\bibliographystyle{unsrt}
\begin{scriptsize}
\bibliography{dissertation-2}
\end{scriptsize}

\newpage
\appendix{\bf\Large Appendices: Algorithmic Details}
\\

In this appendix we present details of inequality formulation for the 
end-to-end performance evaluation. In addition, we present the 
mathematical model to solve these inequalities. We also discuss the case 
of multiple request rounds for the timer suppression mechanism, and present several
example case studies.

\section{Deriving Stress Inequalities}
\label{myapp}

Given the target event, transitions are identified as either
wanted or unwanted transitions, according to the maximization or
minimization objective. For maximization, wanted transitions are
those that establish conditions to trigger the target event,
while unwanted transitions are those that nullify these
conditions.

Let $W$ be the wanted transition, and let $t(W)$ be the time of its
occurrence. Let
$C$ be the condition for the wanted transition, and let $t(C)$ be
the time at which it is satisfied. Let $U$ be the unwanted
transition occurring at $t(U)$.

\begin{figure}[th]
 \begin{center}
  \epsfig{file=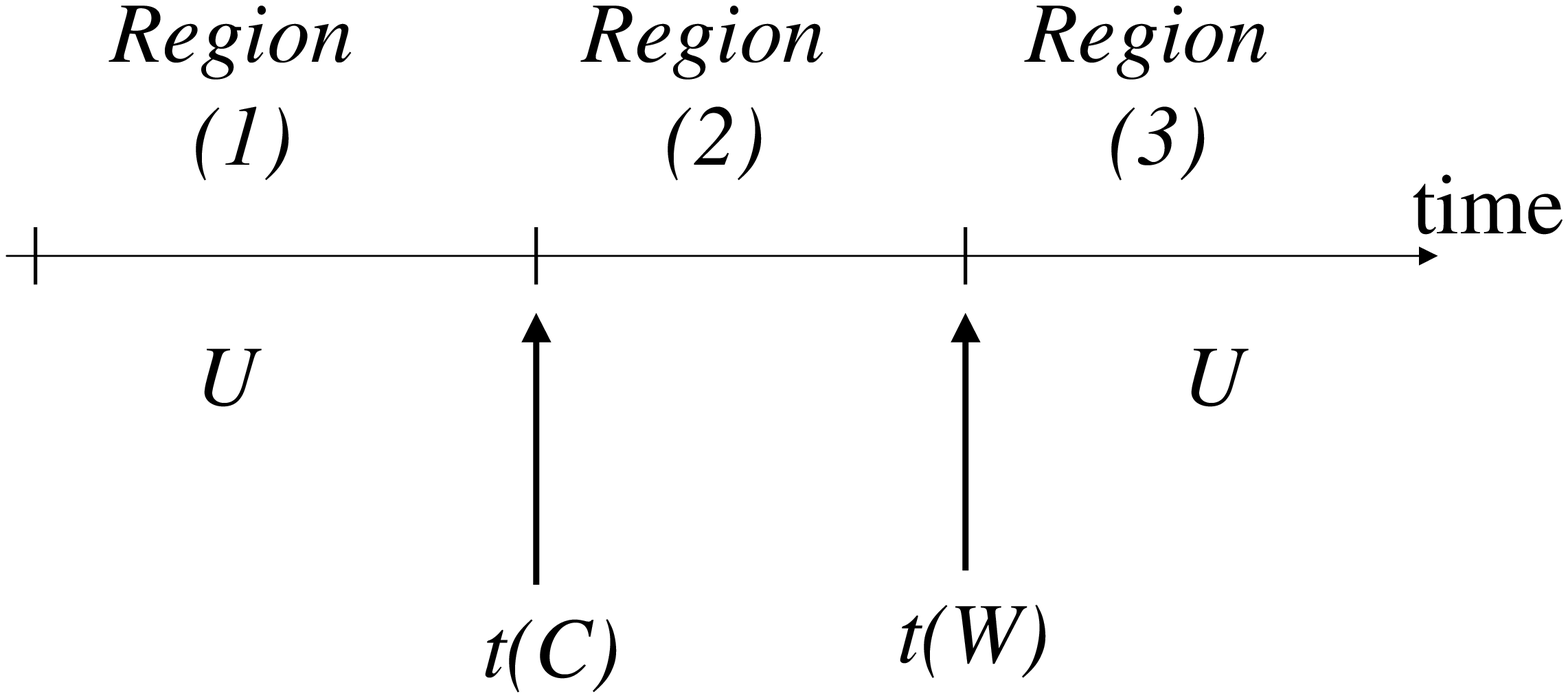,height=5cm,width=6cm,clip=,angle=0}
   \caption{The time-line for transition ordering}\label{timeline}
 \end{center}
\end{figure}

We want to establish and maintain $C$ until $W$ occurs, i.e., in
the duration [$t(C)$, $t(W)$]. Hence, $U$ may only occur outside (before
or after) that interval. In Figure~\ref{timeline}, this means that
$U$ can only occur in $Region (1)$ or $Region (3)$.

Hence, the inequalities must satisfy the following
\begin{enumerate}
\item the condition for the wanted transition, $C$, must be
established before the event for the wanted transition, $W$,
triggers, i.e.,
$t(C) < t(W)$, and
\item one of the following two conditions must be satisfied:
\begin{enumerate}
\item the unwanted transition, $U$, must occur before $C$, i.e.,
$t(U) < t(C)$, or
\item the unwanted transition, $U$, must occur after the wanted
transition, $W$,
i.e., $t(W) < t(U)$.
\end{enumerate}
\end{enumerate}

These conditions must be satisfied for all systems. In addition,
the algorithm needs to verify, using backward search and
implication rules, that no contradiction exists between the above
conditions and the nature of the events of the given protocol.

\subsection{Worst-case Overhead Analysis}

The target event for the overhead analysis is $p_t$. 
The objective for the worst case analysis is to maximize the
number of responses $p_t$. The wanted transition is transition
{\em res\_tmr}, or $Res.(D_T \rightarrow D).p_t$ (see 
Section~\ref{timer}). 
Hence $t(W) = t(p_t)$. The
condition for the wanted transition is $D_T$ and its time is $t(C) = 
t(q_r)$, from
transition {\em tx\_req}, or $q_r.(D \rightarrow D_T)$.

The unwanted transition is one that nullifies the condition
$D_T$. Transition {\em rcv\_res}, or $p_r.(D_T \rightarrow D)$, is 
identified
by the algorithm as the unwanted transition, hence $t(U) =
t(p_r)$.

For a given system $i$, the inequalities become:

\[ t(q_{r_i}) < t(p_{t_i}), \]  
and either 
\[ t(p_{r_{i,j}}) < t(q_{r_i}) \] 
or 
\[t(p_{t_i}) < t(p_{r_{i,j}}). \]

\begin{figure}[t]
 \begin{center}
  \epsfig{file=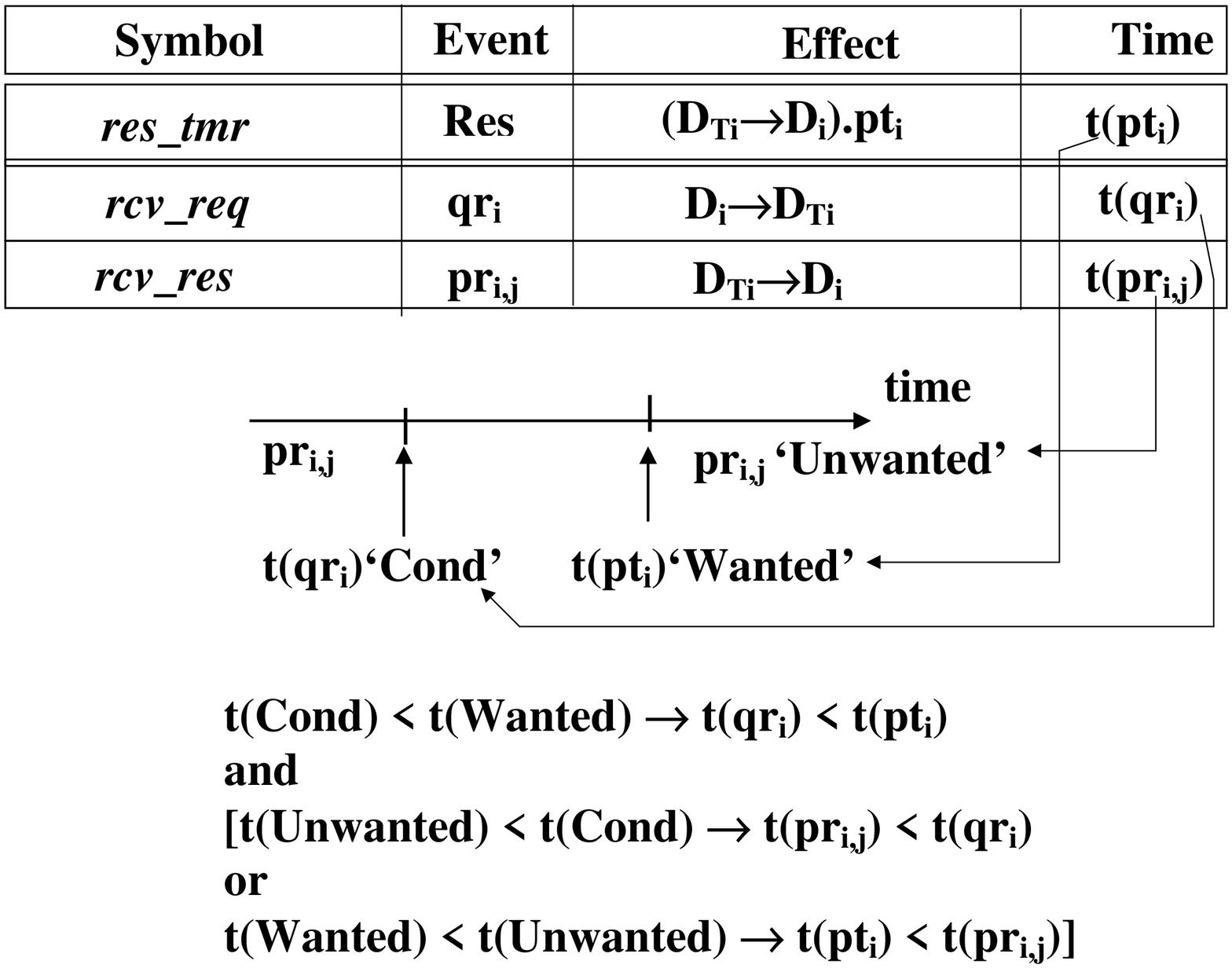,height=6.5cm,width=7.5cm,clip=,angle=0}
   \caption{Formulating the inequalities automatically}\label{table_algo}
 \end{center}
\end{figure}

The above automated process is shown in Figure~\ref{table_algo}.
From the timer expiration implication rule, however, we get that the
response time must have been set earlier by the request
reception, i.e.,
	$Res_i.(D_i \leftarrow D_{T_i}).p_{t_i}
\Leftarrow q_{r_i}.(D_{T_i} \leftarrow D_i)$
and $t(p_{t_i}) = t(q_{r_i}) + Exp_i$. Hence,
$t(q_{r_i}) < t(p_{t_i})$ is readily satisfied and we need not
add any constraints on the expiration timers or delays to
satisfy this condition.
Thus, the inequalities formulated by the algorithm to produce worst-case 
behavior are:

\[ t(p_{r_{i,j}}) < t(q_{r_i}), \]
or
\[ t(p_{t_i}) < t(p_{r_{i,j}}). \]

\subsection{Best-case Analysis}

Using a similar approach to the above analysis, the algorithm 
identifies 
transition {\em rcv\_res}, or $p_r.(D_T \rightarrow D)$, as the wanted
transition. Hence $t(W) = t(p_r)$, and $t(C) = t(q_r)$. The
unwanted transition is transition {\em res\_tmr}, and $t(U) = t(p_t)$.

For system $i$ the inequalities become:

\[ t(q_{r_i}) < t(p_{r_{i,j}}), \] 
and either 
\[ t(p_{t_i}) < t(q_{r_i}) \] 
or 
\[ t(p_{r_{i,j}}) < t(p_{t_i}). \]

But from the backward implication we have $t(q_{r_i}) < t(p_{t_i})$. 
Hence, the algorithm encounters contradiction and the inequality 
$t(p_{t_i}) < t(q_{r_i})$ cannot be satisfied.

Thus, the inequalities formulated by the algorithm to produce best-case 
behavior are:

\[t(q_{r_i}) < t(p_{r_{i,j}}), \] 
and
\[ t(p_{r_{i,j}}) < t(p_{t_i}). \]

\section{Solving the System of Inequalities}
\label{math_model}

In this section we present the general model of the constraints 
(or inequalities) generated by our method. As a first step, we form a 
linear programming problem and attempt to find a solution. If a solution 
is not found, then we form a mixed non-linear programming problem to get 
the maximum number of feasible constraints.

In general, the system of inequalities generated by our method to
obtain worst or best case scenarios, can be formulated as a
linear programming problem.
In our case, satisfying all the constraints, regardless of the
objective function, leads to obtaining the absolute worst/best 
case. For example, in the case of worst case overhead analysis,
this means obtaining the scenario leading to no-suppression.
The formulated inequalities by our method as given in
Section~\ref{overhead} are as follows.
\begin{itemize}
\item for the worst case behavior:
\[ 	d_{Q,i} + Exp_i < d_{Q,j} + Exp_j + d_{j,i}, \]
or
\[ 	d_{Q,i} > d_{Q,j} + Exp_j + d_{j,i}. \]
\item for the best case behavior:
\[ d_{Q,i} + Exp_i > d_{Q,j} + Exp_j + d_{j,i}, \]
and 
\[ 	d_{Q,i} < d_{Q,j} + Exp_j + d_{j,i}. \] 
\end{itemize}

The above systems of inequalities can be represented by a
linear programming model.
The general form of a linear programming (LP) problem is:
\[ Maximize Z = C^TX = \sum_{0\le i\le n} c_i \cdot x_i \]
subject to:
\[ AX \le B \]
\[X \geq 0\]

where $Z$ is the objective function (in our case it is a dummy objective 
function such as $Z = const$), $C$ is a vector of
$n$ constants $c_i$, $X$ is a vector of $n$ variables $x_i$,
$A$ is $m \times n$ matrix, and $B$ is a vector of $m$ elements.
This problem can be solved practically in polynomial time
using Karmarkar~\cite{karmarkar} or simplex method~\cite{simplex}, if a 
feasible solution exists.

In some cases, however, the absolute worst/best case may not be
attainable, and it may not be possible to find a feasible
solution to the above problem. In such cases we want to obtain
the maximum feasible set of constraints in order to get the
worst/best case scenario. To achieve this, we define the problem
as follows:

\[ Maximize \sum_{0\le i\le m} y_i \]
subject to:
\[ y_i \cdot f_i(x) \le 0, \forall i \]
\[y_i \in \{0,1\} \] or 
\[y_i \cdot (1 - y_i) = 0 \]

where $f_i(x)$ is the original constraint from the previous
problem.

This problem is a mixed integer non-linear programming (MINLP)
problem, that can be solved using branch and bound methods~\cite{minlp}.

{\bf Obtaining} {\bf\em Link} {\bf Delays:}

In the previous discussion we assumed that the model deals only with 
end-to-end delays ($d_{i,j}$ of the delay matrix $D$). In some 
cases, 
however, it may be the case that the connectivity of the network topology 
is given and the task is to find the {\em link} delays (instead of 
end-to-end delays). 
We present a very simple extension to the model to accommodate such 
situation, as follows. Let $l_x$ be any link in the topology and let 
$d_{l_x}$ be its delay. Take any two end systems $i$ and $j$ and let the 
path from $i$ to $j$ pass through links $l_a, l_b, \dots \l_n$.
Hence, we get $d_{i,j} = \sum_{x\in L} d_{l_x}$,
where $L = \{l_a, l_b, \dots \l_n\}$. Substituting these relations in the 
above inequalities we can formulate the problem in terms of link delays.

\section{Multiple request rounds}
\label{multi_request}

In Section~\ref{overhead} we conducted the protocol overhead
analysis with the assumption that recovery will occur in one
round of request.
In general, however, loss recovery may require multiple
rounds of request, and we need to consider the request timer as
well as the response timers.
	Considering multiple timers or stimuli adds to the
branching factor of the search. Some of these branches may 
not satisfy the timing and delay constraints. It would be more
efficient then to incorporate timing semantics into the search
technique to prune off infeasible branches.

\begin{figure}[th]
 \begin{center}
  \epsfig{file=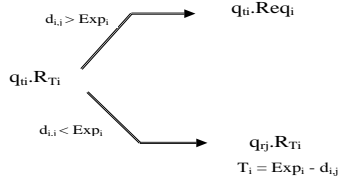,height=5cm,width=8.5cm,clip=,angle=0}
   \caption{Forward search for multiple simultaneous
events}\label{multi_fwd}
 \end{center}
\end{figure}

Let us consider forward search first. For example, consider the
state $q_{t_i}.R_{T_i}$ having a transmitted request
message and a request timer running.
Depending on the timer expiration value $Exp_i$ and the delay
experienced by the message $d_{i,j}$, we may get different
successor states. If $d_{i,j} > Exp_i$ then the request timer
fires first triggering the event $Req_i$ and we get
$q_{t_i}.Req_i$ as the successor state. 
Otherwise, the request message will be received first, and
the successor state will be $q_{r_j}.R_{T_i}$. Note that in this
case the timer value must be decremented by $d_{i,j}$.
This is illustrated in figure~\ref{multi_fwd}. The condition for
branching is given on the arrow of the branch, and the timer
value of $i$ is given by $T_i$.

For backward search, instead of decreasing timer values (as is
done with forward search), timer values are increased, and the
starting point of the search is arbitrary in time, as opposed to
time `0' for forward search.

To illustrate, consider the state having
$(D_i \leftarrow D_{T_i}).R_{T_j}$, with the request timer
running at $j$ and the response timer firing at $i$.

\begin{figure}[th]
 \begin{center}
  \epsfig{file=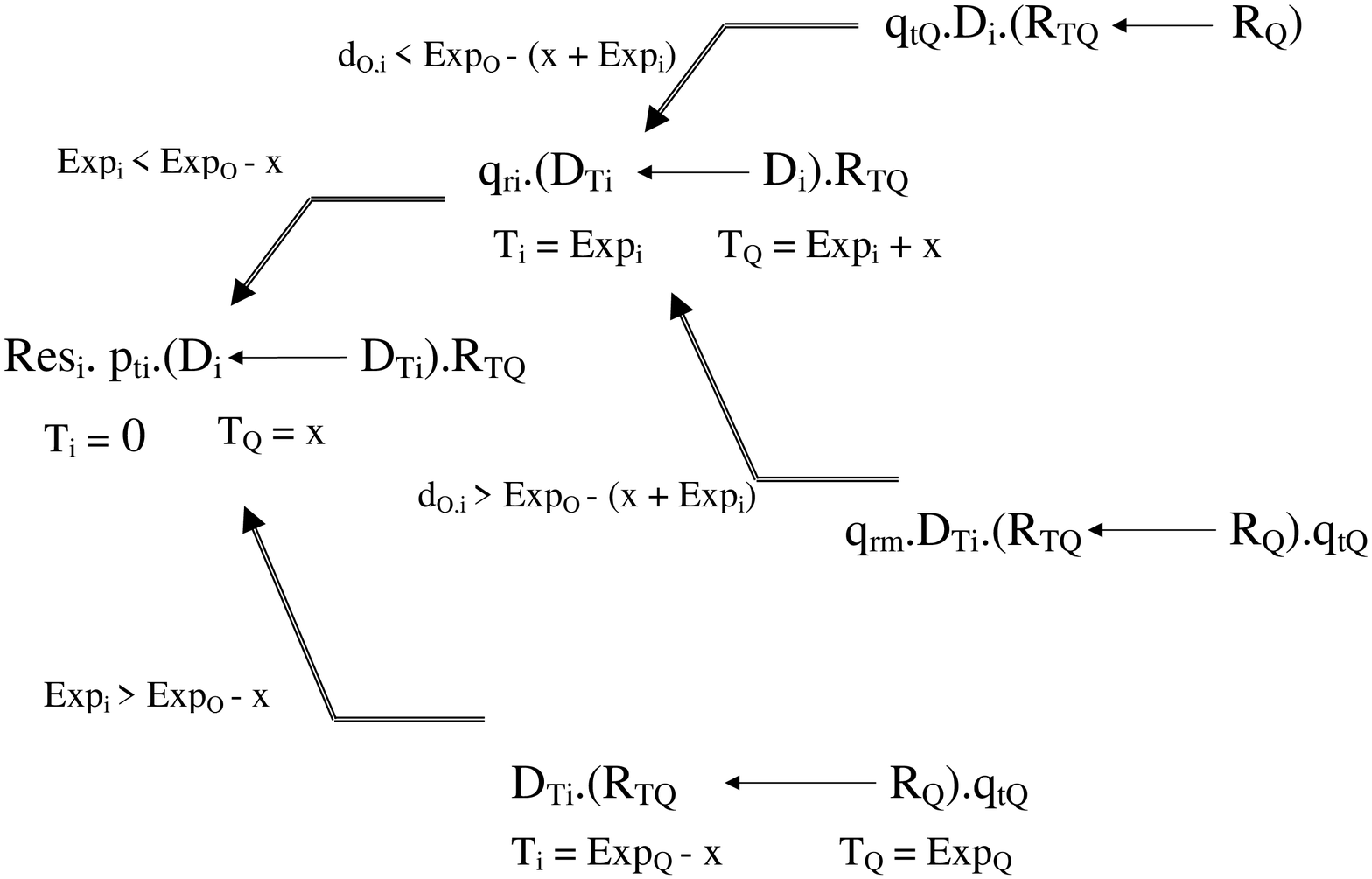,height=5cm,width=7.2cm,clip=,angle=0}
   \caption{Backward search for multiple simultaneous 
events}\label{multi_bkwd}
 \end{center}
\end{figure}

Figure~\ref{multi_bkwd}, shows the backward branching search,
with the timer values at each step and the condition for each
branch. In the first state, the timer $T_Q$ starts at an
arbitrary point in time $x$, and the timer $T_i$ is set to
`0' (i.e. the timer expired triggering a response $p_{t_i}$).
One step backward, either
the timer at $i$ must have been started `$Exp_Q - x$' units in
the past, or the response timer must have been started `$Exp_i$'
units in the past. Depending on the relative values of these
times some branch(es) become valid. The timer values at each step are
updated accordingly. Note that if a timer expires while a message 
is in flight (i.e. transmitted but not yet received), we use the
$m$ subscript to denote it is still multicast, as in $q_{r_m}$ in
the figure.

Sometimes, the values of the timers and the delays are given as 
ranges or intervals. Following we present how branching decision are made 
when comparing intervals.

{\bf Branching decision for intervals} 

In order to conduct the search for multiple stimuli, we
need to check the constraints for each branch. To decide on the branches 
valid for search, we compare values of timers and delays. These values 
are often given as intervals, e.g. $[a,b]$.

Comparison of two intervals $Int_1 = [a_1,b_1]$ and 
$Int_2 = [a_2,b_2]$ is done according to the following rules.

Branch $Int_1 > Int_2$  becomes valid if there exists a value in
$[a_1,b_1]$ that is greater than a value in $[a_2,b_2]$, i.e.
if there is overlap of more than one number between the
intervals. 
We define the `$<$' and `$=$' relations similarly, i.e., if there are any 
numbers in the interval that satisfy the relation then the branch becomes 
valid.

For example, if we have the following branch conditions: 
(i) $Exp_i < Exp_j$, (ii) $Exp_i = Exp_j$, and 
(iii) $Exp_i > Exp_j$.
If $Exp_i = [3,5]$ and $Exp_j = [4,6]$, then, according to our 
above definitions, all the branch conditions are valid. However, if 
$Exp_i = [3,5]$ and $Exp_j = [5,7]$, then only branches (i) and
(ii) are valid.

The above definitions are sufficient to cover the forward search 
branching. However, for backward search branching, we may have an 
arbitrary value $x$ as noted above.

For example, take the state $(D_i \leftarrow D_{T_i}). R_{T_Q}$.
Consider the timer at $Q$, the expiration duration of which is
$Exp_Q$ and the value of which is $x$, and the timer at $i$,
the expiration duration of which is $Exp_i$ and the value of
which is `0', as given in figure~\ref{multi_bkwd}.
Depending on the relevant values of $Exp_i$ and $Exp_Q - x$
the search follows some branch(es). If $Exp_Q = [a_1,b_1]$, then
$x = [0,b_1]$ and $Exp_Q - x = [0,b_1]$.
Hence, we can apply the forward branching rules described earlier by taking
$Exp_Q - x = [0,b_1]$, as follows.
Since $Exp_i = [a_2,b_2]$, where $a_2 >0$ and $b_2 > 0$, hence, the branch
condition $Exp_i > Exp_Q - x$ is always true.
The condition $Exp_i = Exp_Q - x$ is valid when: (i) $Exp_i =
Exp_Q$, or (ii) $Exp_i < Exp_Q$. The last condition, $Exp_i <
Exp_Q - x$, is valid only if $Exp_i < Exp_Q$.

These rules are integrated into the search algorithm for our method to 
deal with multiple stimuli and timers simultaneously.

\section{Example Case Studies}
\label{example}

In this section, we present several case studies that show how to apply 
the previous analysis results to examples in reliable multicast and 
related protocol design problems.

\subsection{Topology Synthesis}

In this subsection we apply the test synthesis
method to the task where the timer values are known and
the topology (i.e., $D$ matrix) is to be synthesized according to
the worst-case behavior. We explore various timer settings. We
investigate two examples of 
topology synthesis, one uses timers with fixed randomization intervals 
and the other uses timers that are a function of distance.

Let $Q$ be the requester and $1$, $2$ and $3$ be potential
responders.
Let 
$V_{t_i}$ be the time
required for system $i$ to trigger a response transmission from 
the time a request was sent, i.e., $V_{t_i} = d_{Q,i} + Exp_i$. 
From Section~\ref{overhead}, we get 
	$V_{t_i} < V_{t_j} + d_{j,i}$ for worst-case overhead.

At time $t_0$ $Q$ sends the request.
For simplicity we assume, without loss of generality, that
the systems are ordered such that $V_{t_i} < V_{t_j}$ for 
$i < j$
(e.g., system $1$ has the least $d_{Q,1} + Exp_1$, then 2, and 
then 3).
Thus the inequalities $V_{t_i} < V_{t_j} + d_{j,i}$ are readily
satisfied for $i < j$ and we need only satisfy it for 
$i > j$.

From equation (1) for the worst-case (see Section~\ref{overhead}) we get:

\begin{eqnarray}
V_{t_2} < V_{t_1} + d_{1,2}, \nonumber\\
V_{t_3} < V_{t_1} + d_{1,3}, \nonumber\\
V_{t_3} < V_{t_2} + d_{2,3}.
\end{eqnarray}

By satisfying these inequalities we obtain the delay settings of the worst
case topology, as will be shown in the rest of this section.

\subsubsection{Timers with fixed randomization intervals}

	Some multicast applications and protocols (such as wb~\cite{SRM},
	IGMP~\cite{igmp} or PIM~\cite{PIM-ARCHv2}) employ fixed
	randomization intervals to set the suppression timers.
	For instance, for the shared white board
	(wb)~\cite{SRM}, the response timer is assigned a random 
	value from the (uniformly distributed) interval
	[t,2*t] where t = 100 msec for the source $src$, and 200
	msec for other responders.

	Assume $Q$ is a receiver with a lost packet. Using wb parameters 
we get $Exp_{src} = [100,200]$ msec, and $Exp_i = [200,400]$
	msec for all other nodes.

To derive worst-case topologies from inequalities (A.1) 
we may use a standard mathematical tool for linear or non-linear 
programming, for more details see Appendix B. 
However, in the following we illustrate general techniques that may be 
used to obtain the solution.

	From inequalities (A.1) we get:

	$d_{Q,2} + Exp_2 = V_{t_2} < V_{t_1} + d_{1,2} = d_{Q,1}
+ Exp_1 + d_{1,2}$.

This can be rewritten as

\begin{equation}
d_{Q,2} - (d_{Q,1} + d_{1,2}) < Exp_1 - Exp_2 =
diff_{1,2},
\end{equation}

where

\footnotesize

\begin{equation}
diff_{1,2} =
  \begin{cases}
     $[100,200] - [200,400] = [-300,0]$ & \text{if 1 is src},\\
     $[200,400] - [100,200] = [0,300]$ & \text{if 2 is src},\\
     $[200,400] - [200,400] = [-200,200]$ &
\text{Otherwise}.
  \end{cases} \notag
\end{equation}

\small


	Similarly, we derive the following from inequalities for 
$V_{t_3}$: 
\begin{center}
	$d_{Q,3} - (d_{Q,1} + d_{1,3}) < diff_{1,3}$, and
	
	$d_{Q,3} - (d_{Q,2} + d_{2,3}) < diff_{2,3}$.
\end{center}
If we assume system 1 to be the source, and for a conservative solution 
we choose the minimum value of $diff$, we get:
\begin{center}
$min(diff_{1,2}) = min(diff_{1,3}) = -300$, 

$min(diff_{2,3}) = -200$.
\end{center}
We then substitute these values in the above inequalities, and
assign the 
values of some of the delays to compute the others. 

{\em Example:}
if we 
assign $d_{Q,1} = d_{Q,2} = d_{Q,3} = 100$msec, we get: 
$d_{1,2} > 300$, $d_{1,3} > 300$ and $d_{2,3} > 200$.

These delays exhibit worst-case
behavior 
for the {\em timer suppression mechanism}.

\subsubsection{Timers as function of distance}

In contrast to fixed timers, this section uses timers that are function
of an estimated distance.
The expiration timer may be set as a function of the
distance to the requester. 
For example, system
$i$ may set its timer to repond to a request from system $Q$ in
the interval:
$[C_1 * E_{i,Q} , (C_1+C_2) * E_{i,Q}]$,
where $E_{i,Q}$ is the estimated distance/delay from $i$ to $Q$,
which is calculated using message exchange (e.g. SRM
session messages) and is equal
to $(d_{i,Q} + d_{Q,i})/2$. (Note that this estimate assumes
symmetry which sometimes is not valid.)

\cite{SRM} suggests values for $C_1$ and $C_2$ as 1 or $log_{10} G$, where
$G$ is the number of members in the group.
	
We take $C_1 = C_2 = 1$ to synthesize the worst-case topology. We get the 
expression

$Exp_1 - Exp_2 = [(d_{1,Q}+d_{Q,1})/2,d_{1,Q}+d_{Q,1}] -
[(d_{2,Q}+d_{Q,2})/2,d_{2,Q}+d_{Q,2}]$.

{\em Example:}
If we assume that
$d_{1,Q}=d_{Q,1}=d_{2,Q}=d_{Q,2}=100msec$, we can rewrite the
above relation as
$Exp_1 - Exp_2 = [-100, 100]$ msec.

Substituting in equation (A.2) above, we get $d_{1,2} > 100$msec.
Under similar assumptions, we can obtain $d_{2,3} > 100$msec, and
$d_{1,3} > 100$msec.

Topologies with the above delay settings will experience the
worst case overhead behavior (as defined above) for the {\em
timer suppression} mechanism.

As was shown, the inequalities formulated automatically by our method in
section~\ref{overhead}, can be used with various timer strategies (e.g.,
fixed timers or timers as function of distance). Although the topologies we
have presented are limited, a mathematical tool (such as LINDO) can be used 
to obtain solutions for larger topologies.

\subsection{Timer configuration}

In this subsection we give simple examples of the timer
configuration task solution, where the delay bounds (i.e., D
matrix) are given and the timer values are adjusted to achieve
the required behavior.

In these examples the delay is given as an interval [x,y] msec. We show 
an example for worst-case analysis.

\subsubsection{Worst-case analysis}

If the given ranges for the delays are [2,200] msec for all
delays, 
then the term $d_{Q,j} - d_{Q,i} + d_{j,i}$ evaluates to
[-196,398]. From equation (A.2) above, we get

$Exp_i < Exp_j - 196$, to guarantee that a response is triggered.

If the delays are [5,50] msec, we get:

\[ Exp_i < Exp_j - 45, \]

i.e., $i$'s expiration timer must be less than $j$'s by at least
45 msecs.
Note that we have an implied inequality that $Exp_i > 0$ for
all $i$.

These timer expiration settings would exhibit worst-case behavior for the
given delay bounds.

%
%
%
%

\end{document}